\documentclass[num-refs]{wiley-article}

\usepackage{amsmath}
\usepackage{multirow}
\usepackage{scalerel}
\usepackage{array}
\usepackage{booktabs}
\usepackage{fixltx2e}

\usepackage{siunitx}
\usepackage{subcaption}

\papertype{Manuscript}
\paperfield{Wind Energy}

\title{Spatio-seasonal risk assessment of upward lightning at tall objects using meteorological reanalysis data}

\author[1]{Isabell Stucke}
\author[1]{Deborah Morgenstern}
\author[1]{Georg J. Mayr}
\author[2]{Thorsten Simon}
\author[3]{Wolfgang Schulz}
\author[3]{Gerhard Diendorfer}
\author[3]{Hannes Pichler}
\author[2]{Achim Zeileis}

\affil[1]{Department of Atmospheric and Cryospheric Sciences, University of Innsbruck, Innsbruck, Tyrol, 6020, Austria}
\affil[2]{Department of Statistics, University of Innsbruck, Innsbruck, Tyrol, 6020, Austria}
\affil[3]{OVE Service GmbH, Department ALDIS, Vienna, 1010, Austria}

\corraddress{Department of Atmospheric and Cryospheric Sciences, University of Innsbruck, Innsbruck, Tyrol, 6020, Austria}
\corremail{isabell.stucke@uibk.ac.at}

\fundinginfo{Austrian Climate Research Program - Implementations, Grant Number: KC305650}

\runningauthor{Stucke et al.}

\begin{document}
\begin{frontmatter}
\maketitle

\begin{abstract}
This study investigates lightning at tall objects and evaluates the risk of upward lightning (UL) over the eastern Alps and its surrounding areas. While uncommon, UL poses a threat, especially to wind turbines, as the long-duration current of UL can cause significant damage. Current risk assessment methods overlook the impact of meteorological conditions, potentially underestimating UL risks. 
Therefore, this study employs random forests, a machine learning technique, to analyze the relationship between UL measured at Gaisberg Tower (Austria) and $35$ larger-scale meteorological variables. Of these, the larger-scale upward velocity, wind speed and direction at 10 meters and cloud physics variables contribute most information.
The random forests predict the risk of UL across the study area at a 1~km$^2$ resolution. Strong near-surface winds combined with upward deflection by elevated terrain increase UL risk. The diurnal cycle of the UL risk as well as high-risk areas shift seasonally. They are concentrated north/northeast of the Alps in winter due to prevailing northerly winds, and expanding southward, impacting northern Italy in the transitional and summer months. 
The model performs best in winter, with the highest predicted UL risk coinciding with observed peaks in measured lightning at tall objects. The highest concentration is north of the Alps, where most wind turbines are located, leading to an increase in overall lightning activity. Comprehensive meteorological information is essential for UL risk assessment, as lightning densities are a poor indicator of lightning at tall objects.

\end{abstract}
\end{frontmatter}
\clearpage
\section{Introduction}
Wind power has become the cornerstone of the transition to a greener and more sustainable future. This transition is being driven by the continued expansion of wind turbines as well as by investments to extend the life time of existing facilities.
The sensitive turbines are exposed not only to the wind that generates the electricity, but also to various other forces of nature. Among these natural forces, lightning has gained particular attention in recent years \cite[e.g.,][]{IEC2019,Candela2016,Montanya2016}. 
Depending on both the physical height of the turbine and its elevation relative to the surrounding terrain,  it can be exposed to a strong amplification of the electric field. This amplification is often expressed in terms of the effective height. The effective height is larger if a tall object is located on a mountain or hill \cite[e.g.,][]{Zhou2010,Shindo2018}.
For objects with effective heights below about 100~m, the main proportion of lightning at tall objects is assumed to be downward lightning (DL). For objects with an effective height greater than 100~m, a critical proportion of lightning can be upward lightning (UL). 
UL only initiates from tall objects and propagates upward towards the charged thundercloud. For objects with effective heights greater than 500 m, all lightning is assumed to be UL \cite{Rakov2003a}. 

Although rare, UL may cause considerable damage to wind turbines. A particularly prolonged current flow can transfer large amounts of charge, which can lead to the melting of individual rotor blades or even the complete failure of the turbine \cite[e.g.,][]{birkl2017}.
The lightning receptors installed at the tip of the Gaisberg Tower in Salzburg (Austria) reveal that, unlike DL, UL is relatively evenly distributed throughout the year, with a slight preference for the colder seasons \cite{Diendorfer2009}.
Better understanding and predicting these rare events, as well as a better risk assessment, is essential for extending the life of individual existing or planned wind turbines, e.g., by equipping them with appropriate lightning protection devices \cite{IEC2019}.

The most serious problem in a spatio-temporal risk assessment is the lack of necessary data. The UL observations at the Gaisberg Tower show that more than 50~\% of UL never appear in the data of conventional lightning location systems (LLS). This is because conventional LLS cannot detect a particular subtype of UL that does not emit an electromagnetic field strong enough to be detectable and consists only of a long duration initial continuous current (ICC) \cite{Diendorfer2015}. The result is a critical underestimation of the actual UL activity and therefore of total lightning at tall objects. As LLS do not distinguish between UL and DL, in the current study lightning at tall objects may include both DL and UL from an effective height $\geq$~100~m.

Current standards to assess the risk of lightning at wind turbines incorporate technical and topographical features, focusing on three key elements. These include the density of lightning strikes per square kilometer annually, the height of the wind turbine represented by its circular collection area (with a radius three times its height), and a specific environmental factor \cite{IEC2019, Rachidi2008, Pineda2018, March2018}. However, challenges arise in this assessment. The local annual lightning density predominantly considers lightning during the convective warm season when they peak annually, largely overlooking lightning during other seasons and particularly UL, which studies suggest pose a significant threat to wind turbines year-round \cite[e.g.,][]{Becerra2018}. Since UL results from complex atmospheric processes acting on different scales, it is crucial to recognize the significant impact of meteorological conditions. Neglecting these factors might lead to a substantial underestimation of the risk posed by lightning at tall objects, particularly by UL.

Investigating the rare and underrated phenomenon using unique UL observations at the Gaisberg Tower in combination with a wide range of globally available atmospheric reanalysis variables using flexible machine learning techniques offers a great opportunity for better risk assessment compared to the current standards. 
Machine learning can not only compensate for the problem of missing data, but also provide meaningful insights, recognize patterns and achieve better predictability.

The study consists of two main steps. In the first step, random forests based on data from the Gaisberg Tower are used to learn which larger-scale meteorological variables are responsible for triggering UL. The tower-trained models are then applied to a larger study area, including Austria, southern and central Germany, Italy, and Switzerland, to obtain high-resolution (~1~km$^2$~) seasonal and annual UL risk maps for the entire area. In order to better understand the predicted risk, the seasonal variations of the most influential larger-scale meteorological variables found at the Gaisberg Tower are investigated.
 LLS-observed lightning at objects (not just at wind turbines) with an effective height $\geq$~100~m are used to verify the resulting risk maps.

\section{Data}\label{sec:data}
The study requires meteorological data, lightning data and a database of all tall objects within a chosen study area comprised of flat, hilly and complex terrain in the eastern Alps (Fig.~\ref{fig:topomap}). Larger-scale reanalysis data (ERA5) with hourly resolution \cite{Hersbach2020} form the basis of all meteorological investigations in this study. In addition, ground-truth lightning current measurements at the Gaisberg Tower in Salzburg \cite[Austria,][]{Diendorfer2009} and LLS data from the European Cooperation for Lightning Detection \cite[EUCLID,][]{Schulz2016} are used. In order to verify the predicted risk at tall objects, different types of tall objects documented by the national aviation safety authorities of Austria, Switzerland, Germany and Italy are employed \cite{italian_obstacles,austrian_obstacles,swiss_obstacles,german_obstacles}. 
The verification period covers three years (2021--2023).

\subsection{Atmospheric reanalysis}
ERA5 is the fifth generation of global climate reanalysis provided by the European Centre for Medium-Range Weather Forecasts (ECMWF). Data are available at hourly resolution and at a spatial resolution of $31$~km horizontally (~$0.25$~$^\circ$~$\times$~$0.25~^\circ$ latitude-longitude grid) and at $137$ levels vertically. 
Given that a precise risk assessment may necessitate a higher resolution than that offered by ERA5, the ERA5 variables are bilinearly interpolated to a $0.01^\circ$ $\times$ $0.01^\circ$ latitude-longitude grid, roughly equivalent to 1 km $\times$ 1 km.
In this study, 35 different variables from ERA5 are used to explain the occurrence of UL. These are either directly available or derived from variables at the surface, on model levels, or integrated vertically.
A complete list of the variable groups and individual variables can be found in the supporting information.

Atmospheric reanalysis data are first used in the modeling step, where each variable is spatially and temporally interpolated to each UL observation at Gaisberg Tower. They are secondly used in the transfer step to the larger study domain shown in Fig.~\ref{fig:topomap}, where each variable is bilinearly interpolated to each $1~$km$^2$ grid cell within the chosen study area in a verification period between 2021 and 2023. 

\subsection{Lightning measurements}



\begin{figure}
\centering
\includegraphics[width=0.7\textwidth]{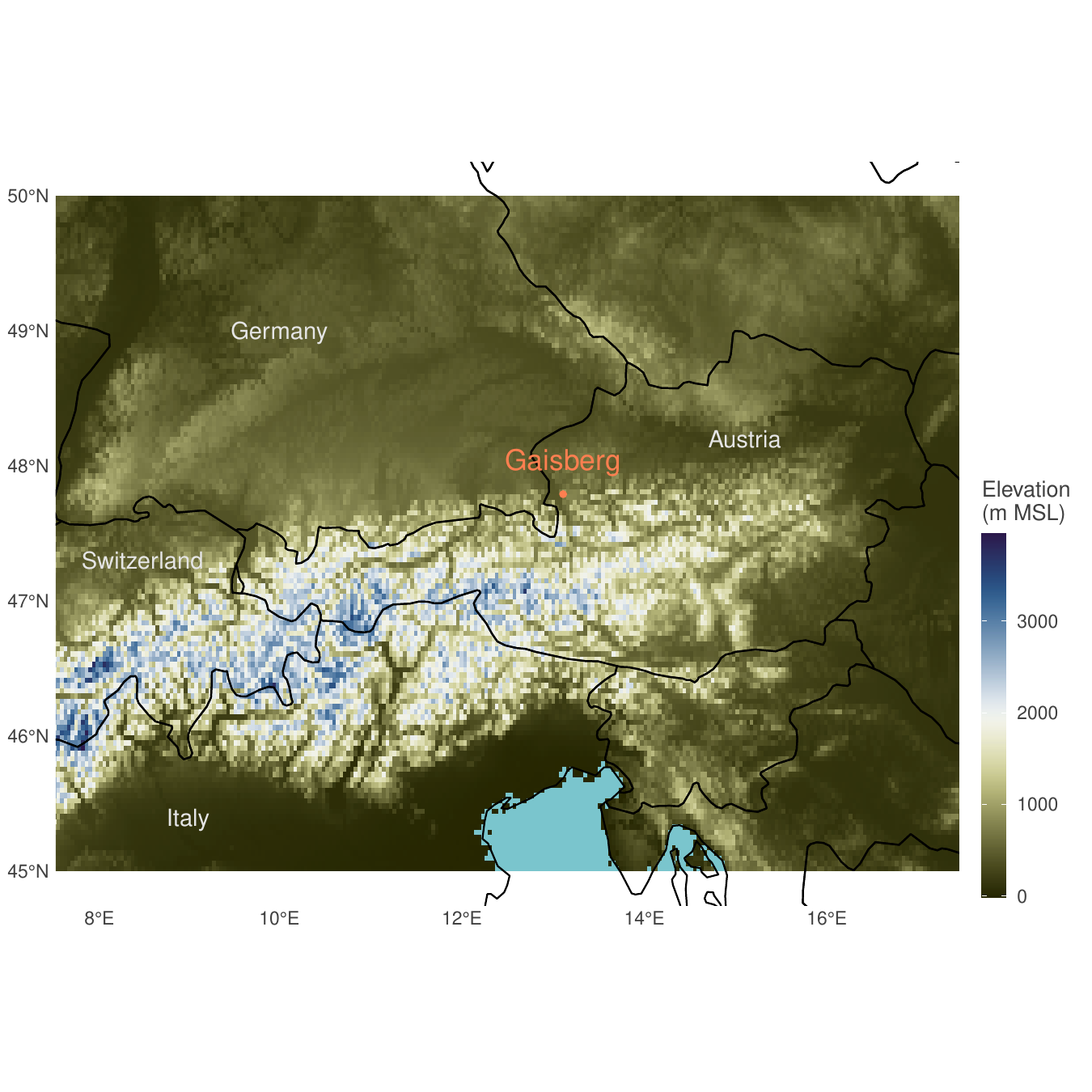}
\caption{Topographic overview of study area and location of the instrumented Gaisberg Tower (Salzburg, Austria). Colors indicates the elevation above mean sea level according to data taken from the Shuttle Radar Topography Mission with a 90 m spatial resolution \cite{farr2000}.}
\label{fig:topomap}
\end{figure}


LLS measurements for the study area (45$^\circ$N--50$^\circ$N and 8$^\circ$E--17$^\circ$E) are from the LLS EUCLID. 
The LLS measures at a frequency range from 400~Hz to 400~kHz and  quantifies lightning flash activity with a median location accuracy of about 100 m  \cite{Schulz2016,diendorfer2016,vergeiner2013}. 
While the LLS detects DL with a detection efficiency of more than 90~\%, the detection efficiency drops to less than 50 \% in the case of UL. Therefore, the proportion of UL can significantly affect the detection efficiency of lightning at tall objects. 

The fundamental data source for constructing models to understand the occurrence of UL is only accessible through direct measurements on specifically instrumented towers.  With a physical height of 100 m above ground and 1,288 m above mean sea level ($47 ^\circ 48'$ N, $13 ^\circ 60'$ E, Fig.~\ref{fig:topomap}), Gaisberg Tower predominantly experiences UL \cite{diendorfer2011}. In total, $956$ UL flashes were recorded at the Gaisberg Tower between 2000 and 2015 and from mid-2020 to the end of 2023.

Equipped with a sensitive shunt-type sensor, Gaisberg Tower measures all UL flashes, irrespective of the current waveform. Three distinct current waveforms are observed at Gaisberg Tower \cite{Diendorfer2009}. The first type emerges when the lightning process ends after the initial phase, involving only a prolonged ICC (ICC\textsubscript{only}). The second type involves this ICC being overlaid with pulse type currents with relative peaks $\geq$~2~kA (ICC\textsubscript{P}). Lastly, the third type of UL evolves after a brief phase of no current followed by one or more downward leader-upward-return stroke processes similar to those observed in DL processes (ICC\textsubscript{RS}).

The measurements at the Gaisberg Tower showed that the ICC\textsubscript{only} subtype cannot be detected by LLS at all. According to \cite{Diendorfer2015}, the other two subtypes of UL presented, (ICC\textsubscript{RS}) and (ICC\textsubscript{P}), are detected by LLS in 96~\% and 58~\% of the cases, respectively.
In order to better verify the resulting models, all analyses in this study are based exclusively on UL that can be detected by LLS, i.e., UL of the ICC\textsubscript{RS} and the ICC\textsubscript{P} type.

\subsection{Lightning at tall objects}

Fortuitously, international aviation regulations require each country to keep and update a database of tall objects that might endanger flight safety.
The study area contains several objects with heights significant for aviation safety (see Table \ref{tab:table1}). This documentation is freely available for Germany, Austria, Switzerland and Italy, but does not include data from the Czech Republic, Slovenia, Hungary and Croatia. The available database gives precise details of the geographic location and physical height of each object, providing a basis for verifying the models from Sect.~\ref{sec:methods1}. Each country is based on a different database with different levels of detail, e.g., tall trees are included in the Swiss database but not in the others. 

UL becomes important only from an effective height of 100~m of the object \cite[e.g.,][]{Rakov2003a}. Hence, the verification process shall extract all LLS-observed lightning that hit an object with an effective height $\geq$~100~m between 2021 and 2023. To match the location accuracy of LLS, all lightning within a radius of 100 meters around each object are considered \cite{diendorfer2016,Soula2019}. 
 
The effective height considers the difference between the height of the object above mean sea level and the height of the surrounding environment. This adjustment to the effective physical height accounts for the electric field enhancement when the mean terrain elevation is significantly lower than the elevation at which an object is located, such as when it is on a mountain or hill. The greater this difference, the greater the effective height and possibly the greater the proportion of total lightning at tall objects.

Several methods have been proposed to compute the effective height. This study uses the method described in \cite{Zhou2010}, which assumes that the mountain is hemispherical with a height equal to the difference between the elevation of where the tall object stands and the average elevation in $1$~km$^2$ around it. The method uses electrical field parameters derived mainly from laboratory experiments. More details are found in \cite{Zhou2010} and in the supplemental information. While this method is readily computable with the information available, it might underestimate the true effective height \cite{Smorgonskiy2012}.

\begin{figure}
\centering
\subfloat{\includegraphics[width=0.5\textwidth]{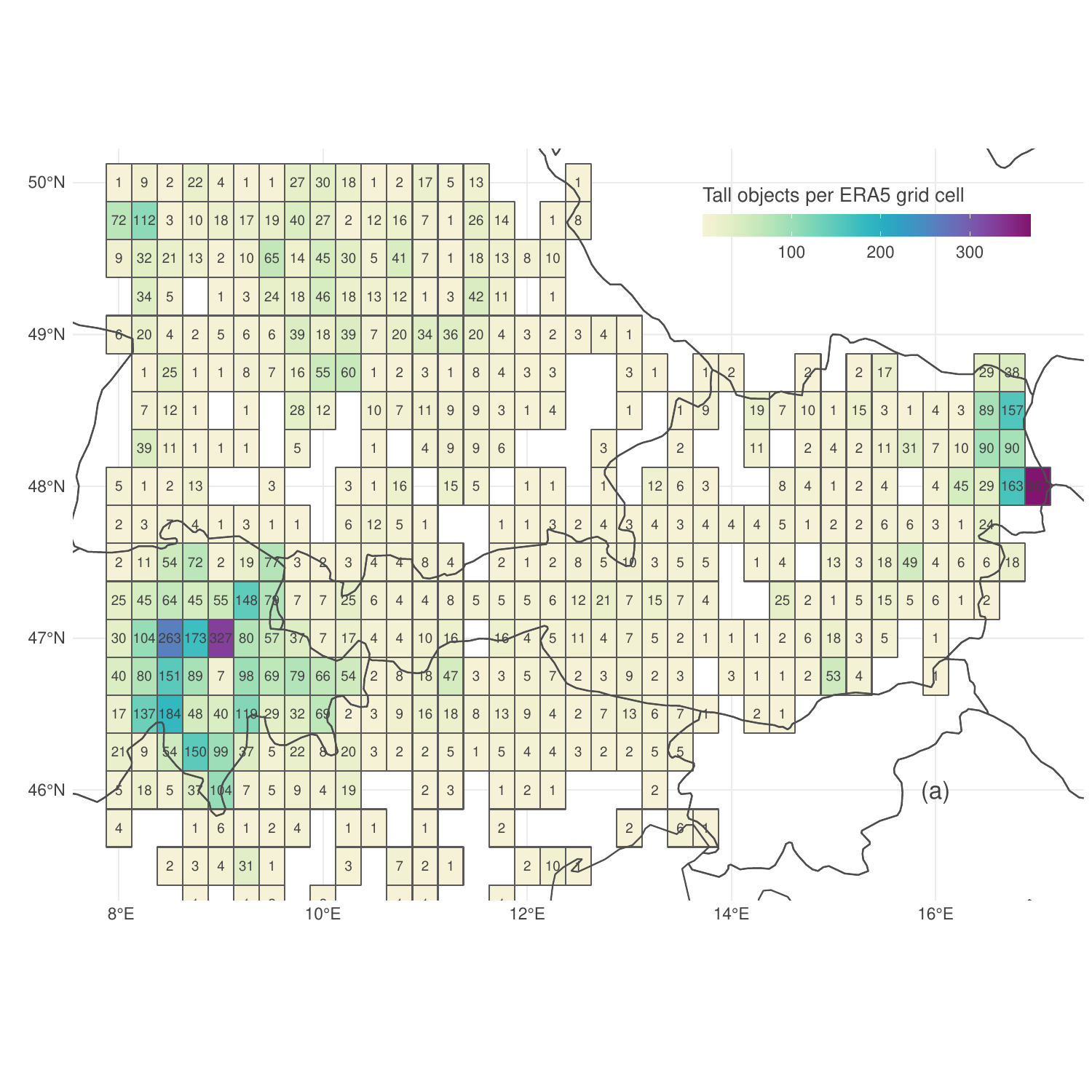}}
\subfloat{\includegraphics[width=0.47\textwidth]{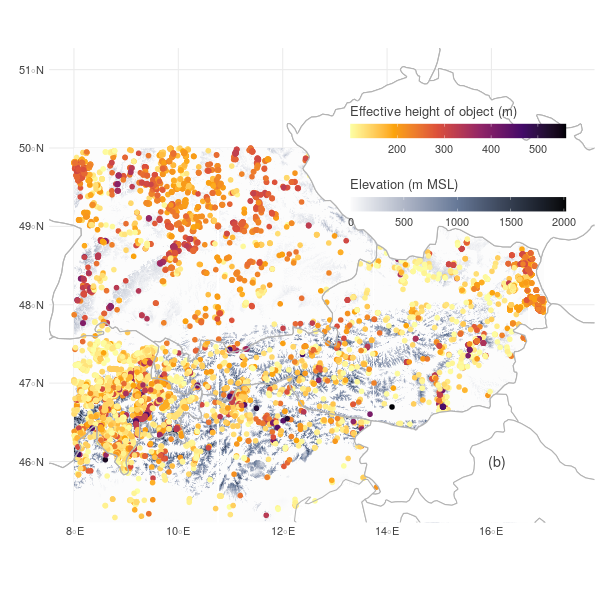}}
\caption{Panel a: accumulated number of objects with effective heights $\geq$100~m in ERA5 grid cells ($0.25^\circ \times 0.25^\circ$). Panel b: all objects with effective heights $\geq$100~m coded by color. }
\label{fig:objects_effheight}
\end{figure}

Figure~\ref{fig:objects_effheight}a gives an overview how tall objects are distributed over the study area and panel b illustrates the distribution of the effective height ($\geq$~100 m) of objects, represented by varying colors. 

The highest concentration of tall objects is observed in the easternmost part of Austria and the central-eastern subarea of Switzerland. There are also some areas in central Germany with an increased number of tall objects.
Interestingly, despite the relatively flat terrain in the German subarea, objects exhibit a comparatively large effective height in contrast to more mountainous terrain (panel b). This phenomenon may be attributed to the hilly terrain in the German subarea. In complex terrain, where mountains dominate the landscape, the mean elevation at the area of 1~km$^2$ is relatively high. Conversely, in hilly terrain, the mean elevation is relatively low, causing hills to stand significantly above the environmental average. 

\begin{table}
\centering
  \caption{List of objects in the national regions of the study area documented by the respective aviation authorities. Listed are the numbers of objects with an effective height $\geq$~100~m and physical height $\geq$~100~m (in parenthesis).}
  \rowcolors{2}{white}{white}
\begin{tabular}{>{\centering\arraybackslash}p{2.5cm}||p{1.3cm}|p{1.3cm}|p{1.3cm}|p{1.2cm}}

  \textbf{Type of object} & \textbf{Austria} & \textbf{German subarea} & \textbf{Italian subarea} & \textbf{Swiss subarea} \\
  \hline \hline
  Wind turbine & 1318 (1283) & 1638 (1632) & 8 (8) & 17 (11) \\
  \hline
  Mast (e.g., antenna, tower) & 270 (26) & 166 (129) & 35 (35) & 90 (12) \\
  \hline
  Building & 35 (35) & 13 (11) & 14 (5) & 25 (5) \\
  \hline
  Stack & 26 (26) & 75 (75) & 30 (30) & 2 (2) \\
  \hline
  Transmission line & 97 (85) & 7 (7) & 75 (75) & 1862 (1216) \\
  \hline
  Cable car & 169 (119) & 1 (1) & 265 (90) & 520 (287) \\
  \hline
  Catenary & 61 (16) & 45 (45) & - & 1169 (566) \\
  \hline
  Others (e.g, vegetation, bridge) & 15 (15) & 12 (3) & 23 (15) & 30 (12) \\
  \hline
  \hline
  Total & 1991 & 1957 & 450 & 3715 \\
  Total per km$^2$ & 0.024 & 0.024 & 0.009 & 0.17\\
  
\end{tabular}

  \label{tab:table1}
\end{table}

\section{Methods}
First, the relationship between UL events and the larger-scale meteorology is analyzed using random forests, linking direct UL measurements from the Gaisberg Tower to meteorological reanalysis data.
 Gaisberg Tower is the only location in the study area where all types of UL are measured.
 The random forests are subsequently applied to the study area and evaluated with LLS-observed lightning at tall objects.

\subsection{Model construction based on Gaisberg Tower data}\label{sec:methods1} 

To link meteorological reanalysis data with the occurrence of UL at the Gaisberg Tower, this study uses random forests, which is a flexible machine learning technique able to tackle nonlinear effects \cite{Breiman2001}.

Whether or not UL occurs at Gaisberg Tower is a binary classification problem. In this classification problem, $35$ larger-scale meteorological variables are the predictors chosen to explain the response. The response is LLS-detectable UL at Gaisberg Tower (1) or no (LLS-detectable) UL (0) at Gaisberg Tower. Each of the meteorological variables is spatio-temporally interpolated to an UL observation at Gaisberg Tower. Excluding LLS undetectable UL (ICC\textsubscript{only}), 549 UL observations are recorded at Gaisberg Tower.

The algorithm constructs decision trees by assessing the connection between the binary response and each predictor variable through permutation tests, also known as conditional inference \cite{Strasser99}. At each recursive step of tree construction, the predictor variable exhibiting the highest (most significant) association with the response variable is chosen. Subsequently, the dataset is partitioned based on this selected predictor variable to optimize the separation of different response classes. This splitting procedure is recursively applied within each subset of the data until a predefined stopping criterion, such as significance or subsample size, is satisfied. A qualitative example of a single decision tree is given in the supporting information.

 In the final stage, the random forest aggregates predictions from this ensemble of trees, thereby enhancing prediction stability and performance. For additional insights into the algorithm and its implementation, refer to \cite{Hothorn2006} and \cite{Hothorn2015}.

The models' response, which indicates the rare presence (1) or very frequent absence (0) of UL, is sampled equally to ensure a balanced representation of the two classes. Hence, the predicted probabilities of the random forest models shown in this study are termed ``conditional probability'' due to the balanced setup of the model response.  To increase the robustness of the results, $10$ different random forest models are used to compute the conditional probability. Each of these random forest models consists of the 549 UL observations associated with the larger-scale meteorological setting and 549 randomly selected non-UL situations. The results shown in this study are the median of these 10 random forests.


\subsection{Transfer of the Gaisberg model result to the study area}

Previous studies by the authors have shown that the random forest models trained on the Gaisberg Tower perform well when tested on withheld data from the Gaisberg Tower or when tested on another tower, the S\"antis Tower in Switzerland \cite[e.g.,][]{stuckejgr}. In this study, the results from the Gaisberg Tower are transferred to a variety of topographic environments from flat to hilly to complex terrain.
The tower-trained random forest model computes the conditional probability of UL in grid cells of 1~km$^2$ and 1 hour from the larger-scale meteorological reanalysis data. 
Whether the resulting models are reasonable is justified by comparing the predicted conditional probabilities with LLS-observed lightning at tall objects as described in Sect.~\ref{sec:data}. 


\section{Results}
The results of the study are presented in three distinct parts.
In order to take into account the factors that critically influence lightning at wind turbines according to the current lightning protection standards, the LLS-observed lightning at tall objects is compared with the total lightning activity including DL to ground within the selected study area (Sect.~\ref{sec:explore}). Then the influence of the effective height of the objects on the LLS-observed lightning is investigated. The section then proceeds to showcase the application of Gaisberg Tower-trained models to the different subareas, illustrating the modeled risk of UL at objects annually and for each season (see Sect.~\ref{sec:risk}). Along with this, the seasonal variations of the modeled risk (Sect.~\ref{sec:riska}) as well as the seasonal variation in the diurnal cycle of the modeled risk is presented (Sect.~\ref{sec:riskb}). Sect.~\ref{sec:riskc} examines the performance of the results by quantitatively comparing the modeled outcomes with LLS-observed lightning at tall objects.
Following this, Sect.~\ref{sec:varimp} investigates the meteorological conditions that predominantly contribute to UL at the Gaisberg Tower. Section~\ref{sec:meteorology} explains the resulting modeled risk from the most important meteorological variables that affect UL risk, including how these influential variables vary throughout the seasons. A case study is included to demonstrate the models' predictive behavior and the conditions leading to an increased risk of UL (Sect.~\ref{sec:casestudy}).

\subsection{LLS-observed lightning at tall objects}\label{sec:explore}

\begin{figure}
\centering
\subfloat{\includegraphics[width=1.06\textwidth]{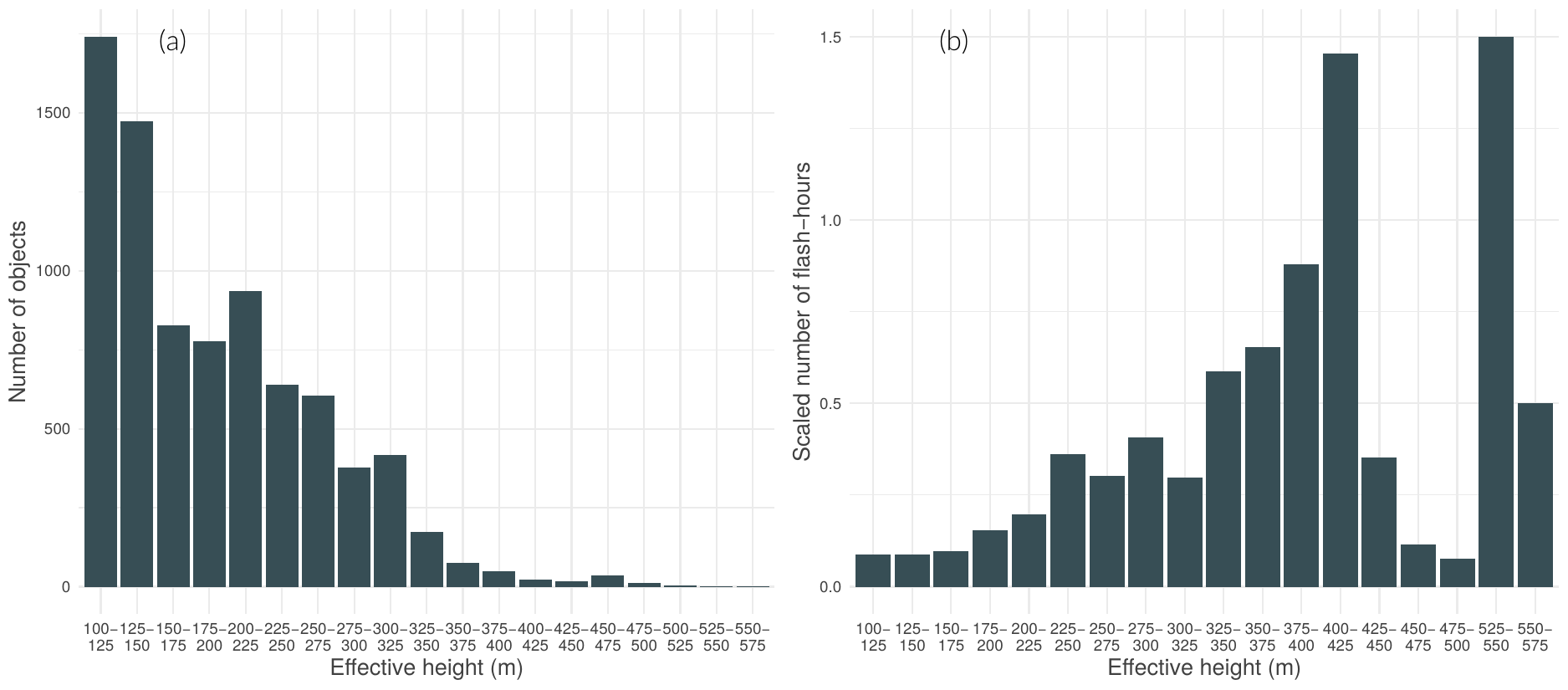}}
\caption{Panel a: number of objects per effective height range. Panel b: number of flash-hours scaled by the number of objects per effective height range.}
\label{fig:effH_statistics}
\end{figure}

As mentioned, current lightning protection standards \cite{IEC2019} take (i) the physical properties of the structure and (ii) the local annual lightning flash density into account.
Considering that the effective height may influence lightning at a tall object according to the standards, panels a and b in Fig.~\ref{fig:effH_statistics} examine the role of effective height on the number of flash-hours for objects with corresponding effective height values.

\begin{figure}
\centering
\subfloat{\includegraphics[width=0.48\textwidth]{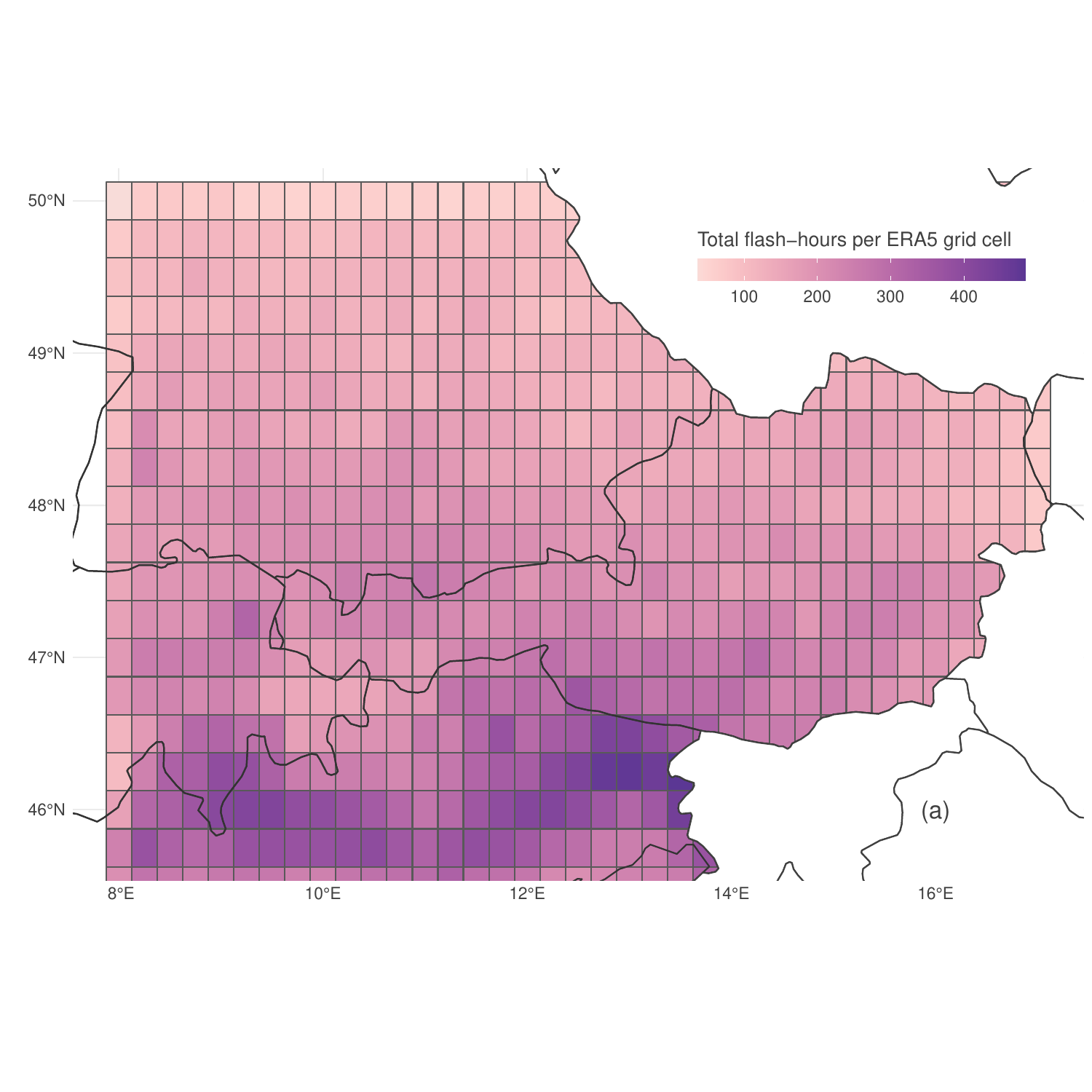}}
\subfloat{\includegraphics[width=0.48\textwidth]{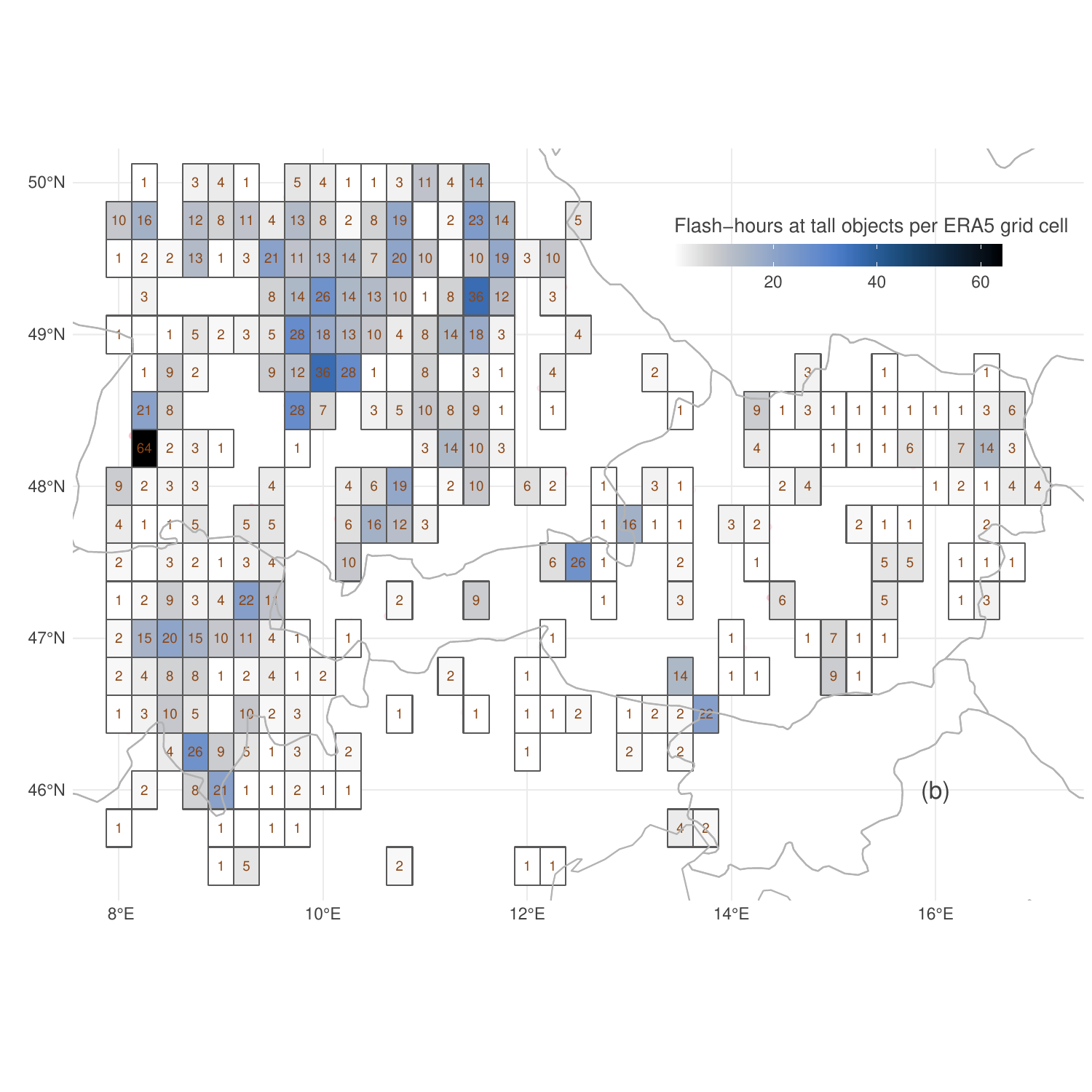}}
\hspace{2cm}
\vspace{-20ex}
\subfloat{\includegraphics[width=0.48\textwidth]{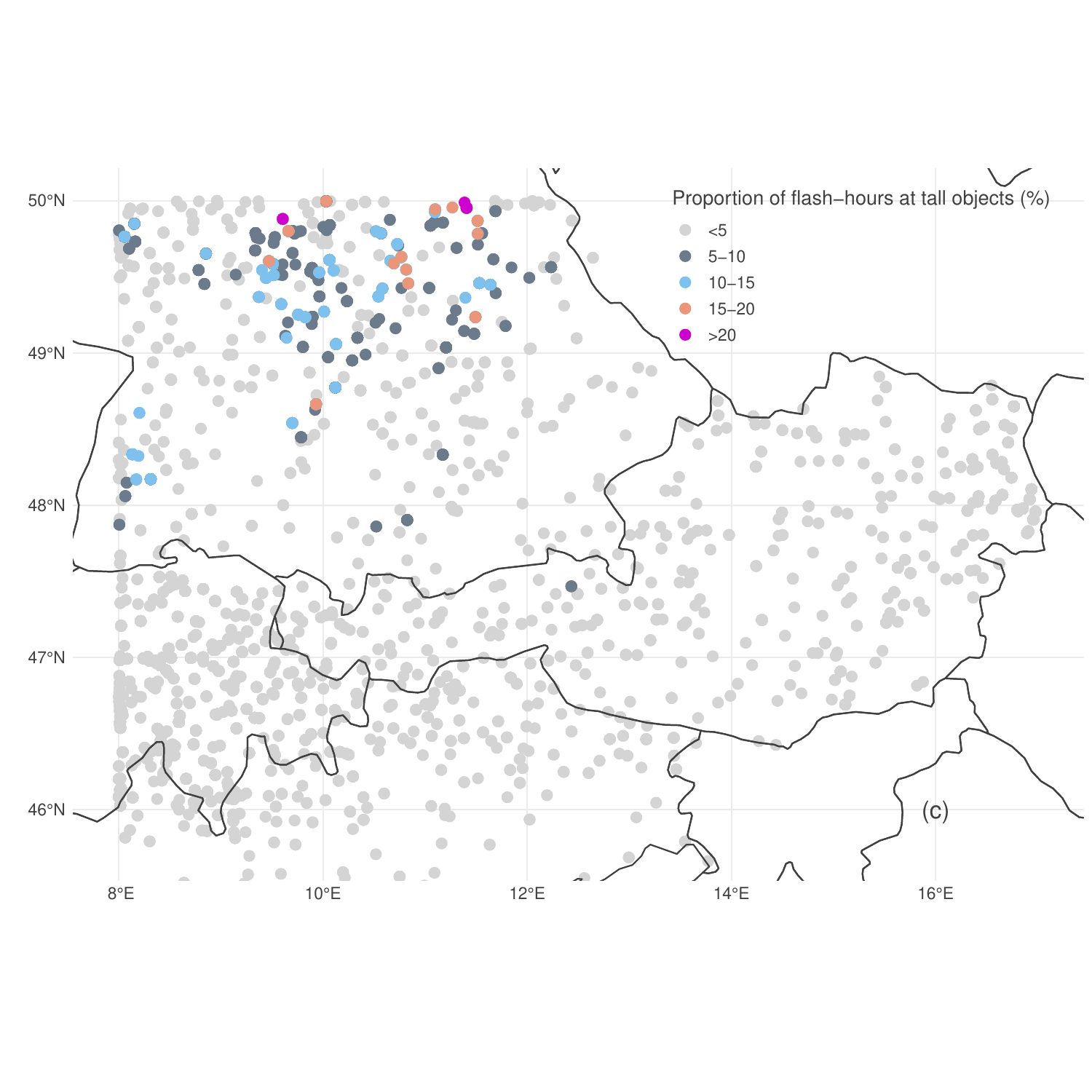}}
\subfloat{\includegraphics[width=0.48\textwidth]{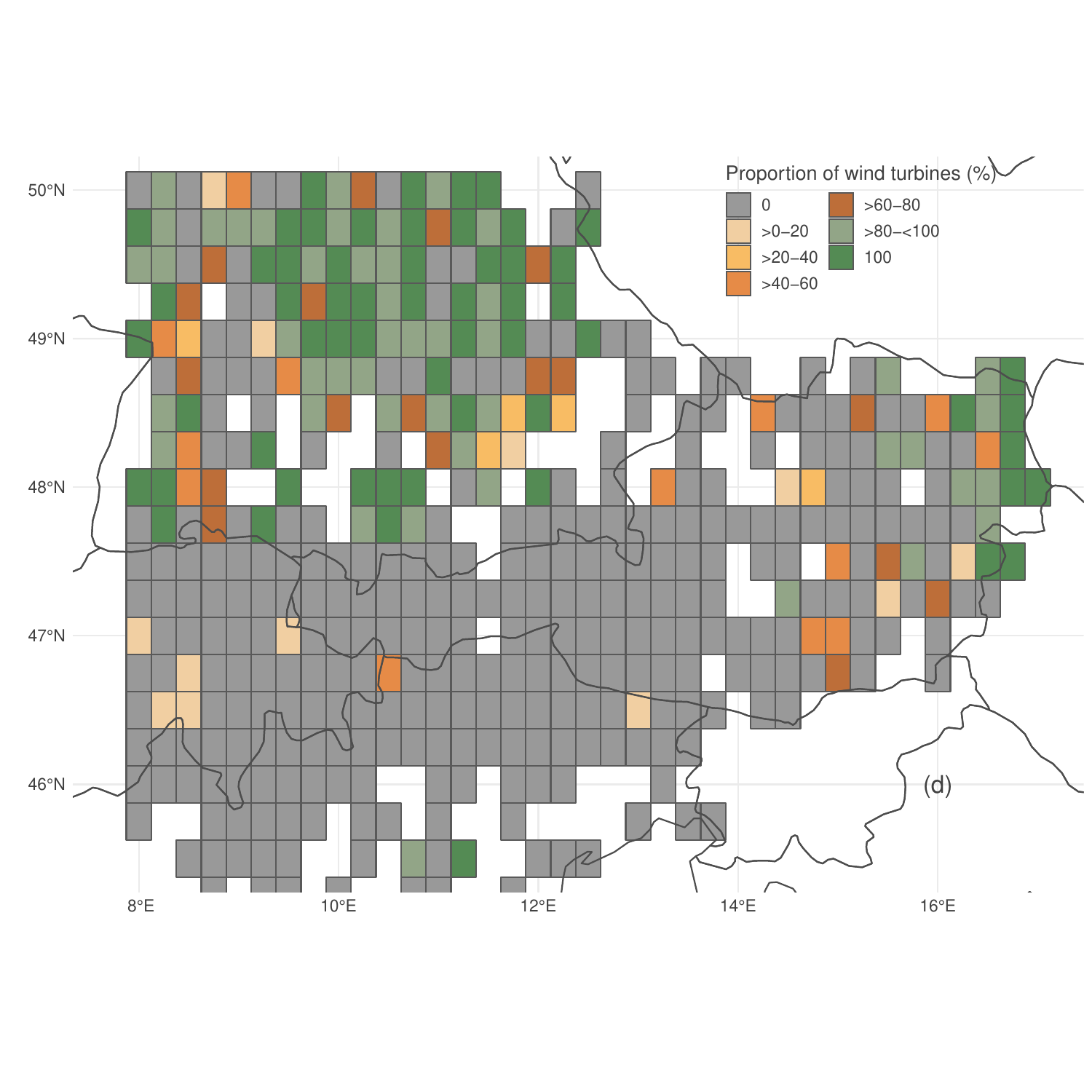}}
\vspace{15ex}
\caption{Panel a: total number of flash-hours in ERA5 grid cell (including DL to the ground and lightning at tall objects) between 2021 and 2023. Panel b: accumulated number of flash-hours at objects with effective heights $\geq$100~m. Panel c: proportion of hours exclusively having lightning at tall objects to the total flash-hours 10~km around each object. Excluded are those flash-hours, where also DL to the ground occurred around the object. Panel d: proportion of wind turbines to the total number of objects in cell. One flash-hour is defined by at least one lightning flash within a grid cell and within one hour.}
\label{fig:explorative}
\end{figure}

Panel a shows that the majority of objects have an effective height around 100~m. 
Panel b shows that objects with higher effective heights are more frequently struck by lightning corroborating previous findings \cite[e.g.,][]{Rakov2003a,Shindo2018}. The gap between 425~m and 500~m is likely due to the very few objects in that height range being located in areas with low overall LLS-observed lightning at tall objects (see Fig.~\ref{fig:explorative}b).
The Gaisberg Tower as computed using the method in \cite{Zhou2010} is in a range between 250 m and 275 m.

The second important factor in assessing the risk of lightning at wind turbines according to the standards is the local annual flash density (Fig.~\ref{fig:explorative}a).
 

Fig.~\ref{fig:explorative}a shows that the highest concentration of the total lightning activity is in the southern part of the study area in northern Italy. These hotspots are thought to result from enhanced moisture transport from the Adriatic Sea by the mountain plain circulation, which hits the rising topography and initiates convection. This is consistent with previous studies investigating lightning climatologies in these regions \cite[e.g.,][]{Simon2022,feudale2013,taszarek2019}. 
 

However, panel b in Fig.~\ref{fig:explorative} is in stark contrast to panel a, as the maximum cumulative flash-hours of lightning at tall objects are concentrated in the southwesternmost part of the German subarea and the central region of the same subarea. In addition, the central-eastern and southernmost parts of Switzerland show a significant accumulation of flash-hours. 
Similarly, panel b in Fig.~\ref{fig:explorative} shows no association with the distribution of objects over the study area in panel a of Fig.~\ref{fig:objects_effheight}.
 
Flash-hours in panel b may have DL to ground in addition to lightning at tall objects within the same hour. To examine the proportion of flash-hours exclusively characterized by lightning at tall objects, panel c examines lightning within a 10 km radius of each object. The panel shows that the high concentration of lightning at tall objects in the Swiss subarea is largely associated with DL to the ground also occurring within 10~km of the tall object within the same hour. In the German subarea, however, the proportion of flash-hours at tall objects with no other lightning activity in the vicinity is significantly higher than in the other subareas. While in most cases hours with exclusively lightning at tall objects accounts for less than 5~\% of the total lightning activity around a tall object, in the German subarea hours with exclusively lightning at tall objects accounts for up to 20~\% or more of the total. It can be assumed that the mere presence of the tall object significantly increases the total lightning activity. From Fig.~\ref{fig:explorative}d it can be concluded that lightning at wind turbines accounts for the largest proportion of lightning activity 10~km around an object in this area, while lightning at wind turbines in the eastern part of Austria, where also many wind turbines are located, accounts for less than 5\% of the surrounding lightning activity.

From this analysis it can be suggested that the local flash density does not sufficiently account for the occurrence of lightning at tall objects and in particular for the occurrence of UL, so that for a more reliable risk assessment detailed meteorological information must be included.

\subsection{Modeled risk of UL at tall objects}\label{sec:risk}
The following analyses highlight the importance of considering the larger-scale meteorological environment for accurate UL risk prediction. The figures show the seasonal variation of the UL risk over the study area as well as the seasonal variation of the diurnal cycle of the UL risk. In addition, the predictive performance of the models is presented and examined seasonally. 

\subsubsection{Seasonal variations of the modeled risk}\label{sec:riska}

\begin{figure}
\centering
\vspace{-30ex}
\subfloat{\includegraphics[width=1\textwidth]{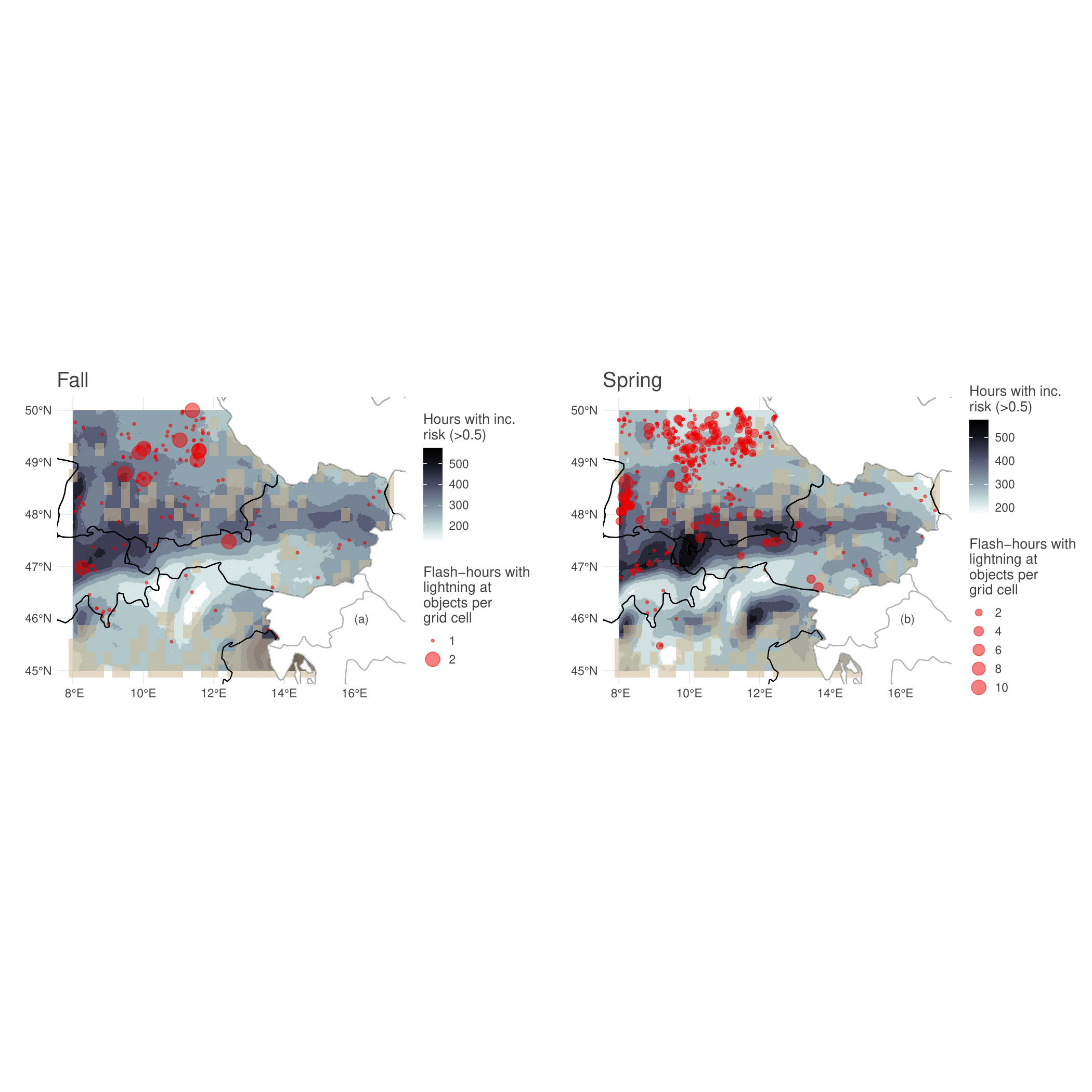}}\vspace{-65ex}
\hspace{4cm}
\subfloat{\includegraphics[width=1\textwidth]{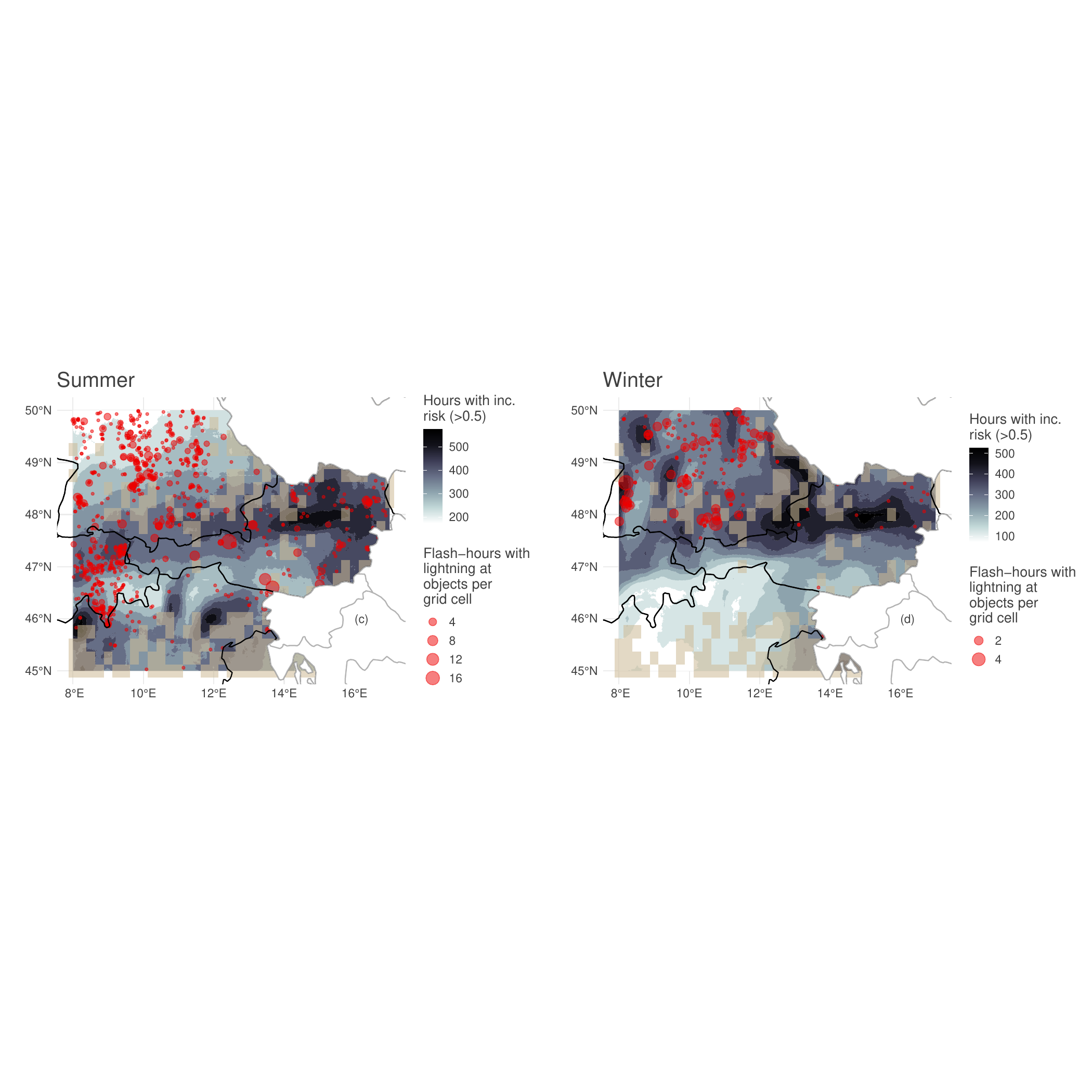}}\vspace{-35ex}
\hspace{20ex}
\subfloat{\includegraphics[width=.55\textwidth]{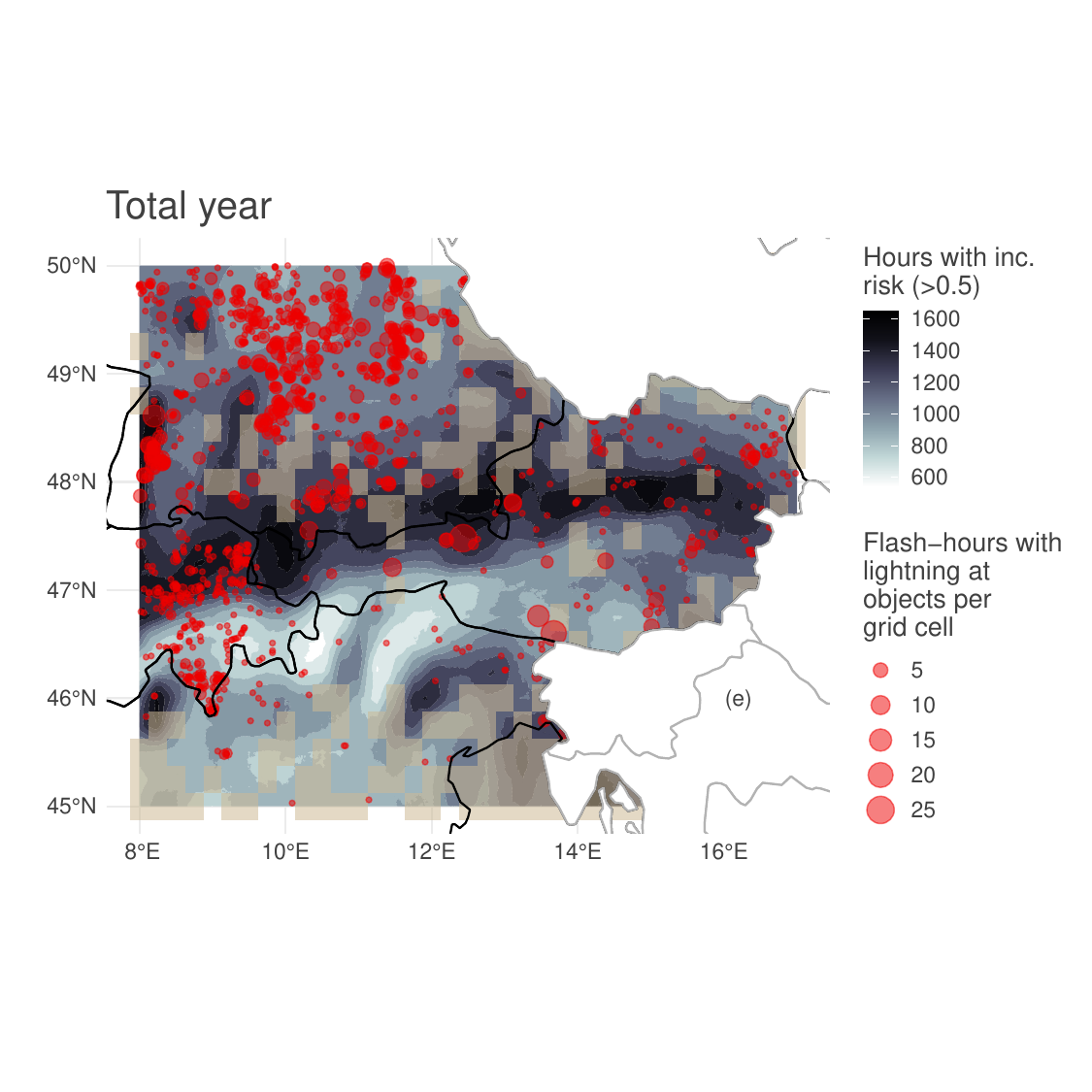}}\vspace{-10ex}
\caption{Seasonal (panels a--d) and annual (panel e) UL risk at tall objects modeled by the Gaisberg Tower-trained random forest models. Risk is quantified by counting the number of hours exceeding a conditional probability of 0.5. Red dots are LLS-detected flash-hours at tall objects accumulated to the 1~km$^2$ grid cell size. The size category numbers are the upper limit, e.g., size category 5 includes flash-hours from 1 to 5. Light beige shaded cells are cells without tall objects.}
\label{fig:seas_risk}
\end{figure}

Panels a--d in Fig.~\ref{fig:seas_risk} depict the risk for fall, spring, summer and winter, while panel (e) presents the annual risk.
Across all five panels, notable regions exhibit increased or decreased risk of UL according to the larger-scale meteorological setting, and these patterns shift with the seasons. Shown is the modeled seasonal (panels a-d) and annual (panel e) risk of UL as predicted by the Gaisberg Tower trained random forests, which are solely based on UL and not DL. Risk is quantified by counting the number of hours in which the models predict a conditional probability greater than 0.5 for each $1$~km$^2$ grid cell. 
Absolute values of increased risk are difficult to interpret because the tower-trained random forests, based on a balanced response with UL and no-UL situations, model the conditional probability. 

The areas with the highest risk of UL shift throughout the year.
From winter through spring and into summer, the areas of increased risk tend to move both southward and eastward. 
In the fall, the region with the highest risk is mainly located in the western German subarea and the southern German subarea, extending into the Swiss and Austrian northern subareas. While similar in spring, there is a slight southward and eastward shift, with the highest risk observed in the westernmost part of Austria extending eastward through Austria along the Alps, the easternmost part of Switzerland, and the southwestern part of Germany. In summer, the hotspot regions shift to the eastern and western parts of northern Italy and the eastern part of Austria. Conversely, in winter, the highest risk extends over most of the German subarea and the northern parts of Switzerland and Austria. In contrast, a rather low risk is observed south of the Alps during the cold season.

Combining the seasonal data reveals a distinct annual pattern (panel e). Areas with a consistently higher risk include the German subarea, the northern parts of Switzerland and north western and central Austria, along with the western and eastern parts of northern Italy.

Looking at LLS-observed lightning at tall objects possibly including DL at tall objects and UL (red dots), it is important to note that more than half of the actual UL flashes may not have been recorded by LLS, as discussed in the introduction. Notably, in winter and the transitional seasons, observed lightning at tall objects is confined to the northern part of the study area, where the highest risk is identified. In contrast, during summer, observed lightning at tall objects extends to the southern regions, where the risk is also increased. 

\subsubsection{Seasonal variations in the diurnal cycle of the modeled risk}\label{sec:riskb}
\begin{figure}
\centering
\includegraphics[width=.8\textwidth]{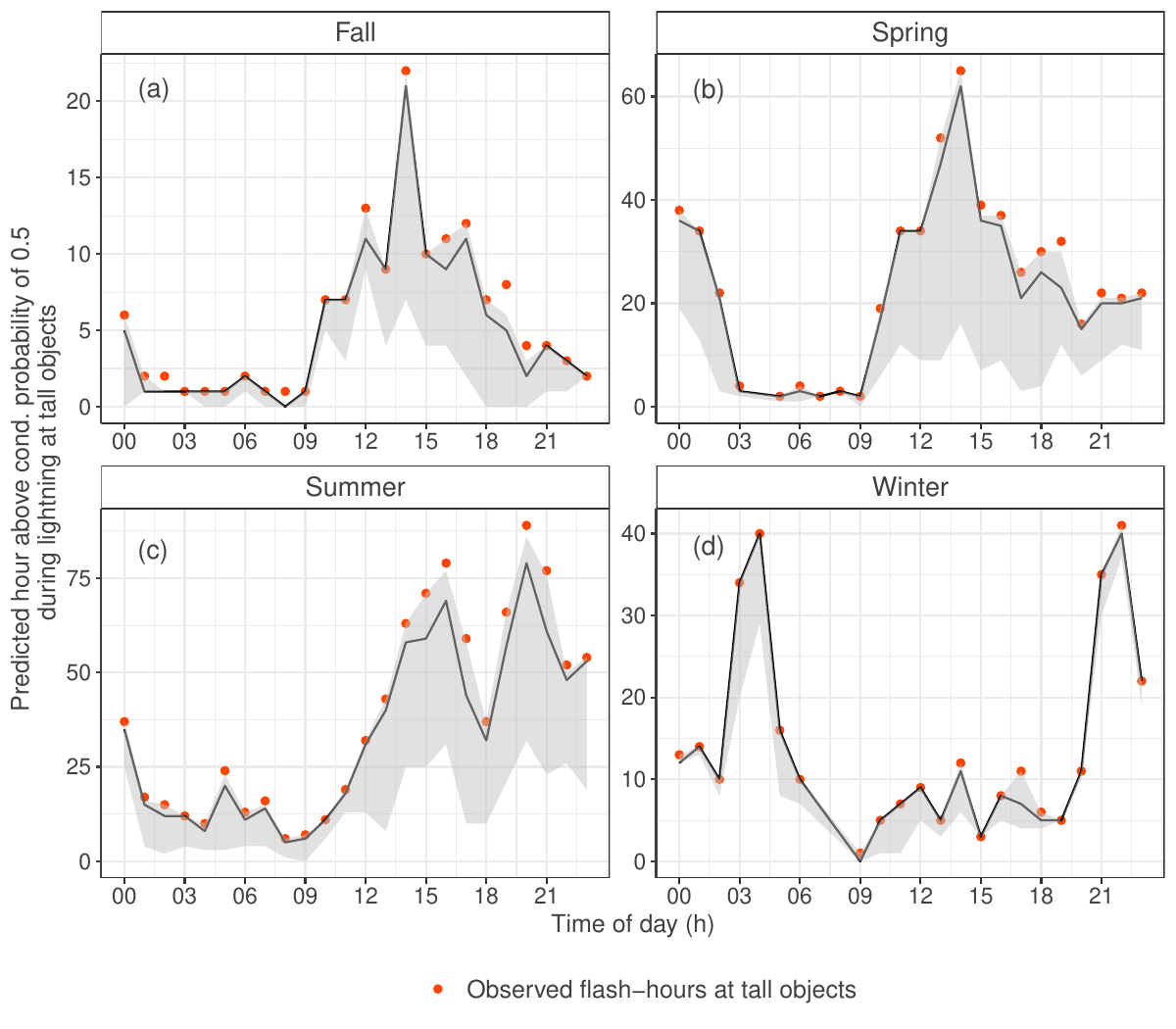}\hspace{3cm}
\caption{Diurnal cycle of accumulated observed flash-hours at tall objects over the entire study area and verification period (orange dots) versus modeled risk of UL during these events (above conditional probability threshold of 0.5, gray line) of UL. The database consists of LLS-observed lightning at tall objects only and neglects situations without lightning at tall objects. As only hourly predictions are provided, situations in which the same object is hit multiple times within the same hour are only counted once. Shaded area shows the difference of the sum of predicted hours between conditional probabilities of 0.25 and 0.75. Smaller shaded areas indicate sharper predictions for identifying lightning at tall objects. The median values in the predictions for UL at tall objects in winter, summer, fall and spring are $0.834$, $0.68$, $0.68$ and $0.67$, respectively.}
\label{fig:daily_cycle}
\end{figure}

Figure~\ref{fig:daily_cycle} panels a--d illustrates that not only does lightning at tall objects vary seasonally, but it also exhibits distinct daily patterns for each season. 

Notably, despite the common substantial increase in DL activity during the summer season, the absolute number of flash-hours at tall objects does not vary as much between seasons as one might expect. The transitional seasons each have a single peak. Activity peaks both in the fall and spring around 14~UTC. The most notable difference between fall and spring is the relatively high activity around midnight in spring, a pattern also observed in summer. Both the summer and winter seasons have two prominent peaks. In summer, the first and second peaks occur around 16~UTC and 19~UTC, respectively, while in winter these peaks occur around 4~UTC and 22~UTC, respectively. This suggests that different meteorological settings may contribute to lightning at tall objects in different seasons, with strong diurnal heating possibly dominating in summer, triggering deep convection and other processes, such as those associated with cold fronts, influencing lightning at tall objects in winter and transitional seasons.

The shaded regions in each panel represent the disparity between aggregating hours with conditional probabilities above 0.25 and those exceeding 0.75. A smaller shaded area indicates sharper \citep{gneiting2007} predictions during observed lightning at tall objects. Contrarily, larger shaded areas indicate that the models barely predicted a conditional probability above 0.75 when lightning was observed at tall objects, indicating less sharpness in the predictions. Among the four seasons, the predictions in winter are sharpest with the most narrow shaded areas particularly during nighttime starting from 20~UTC until around 3~UTC. As the random forests model only UL, the best performance in winter might suggest a greater contribution of UL to all lightning at tall objects in the colder season. Contrarily, the underestimation of random forest models in summer suggests the dominance of DL in lightning at tall objects which the random forest does not account for.


\subsubsection{Model evaluation}\label{sec:riskc}

\begin{figure}
\centering
\subfloat{\includegraphics[width=0.5\textwidth]{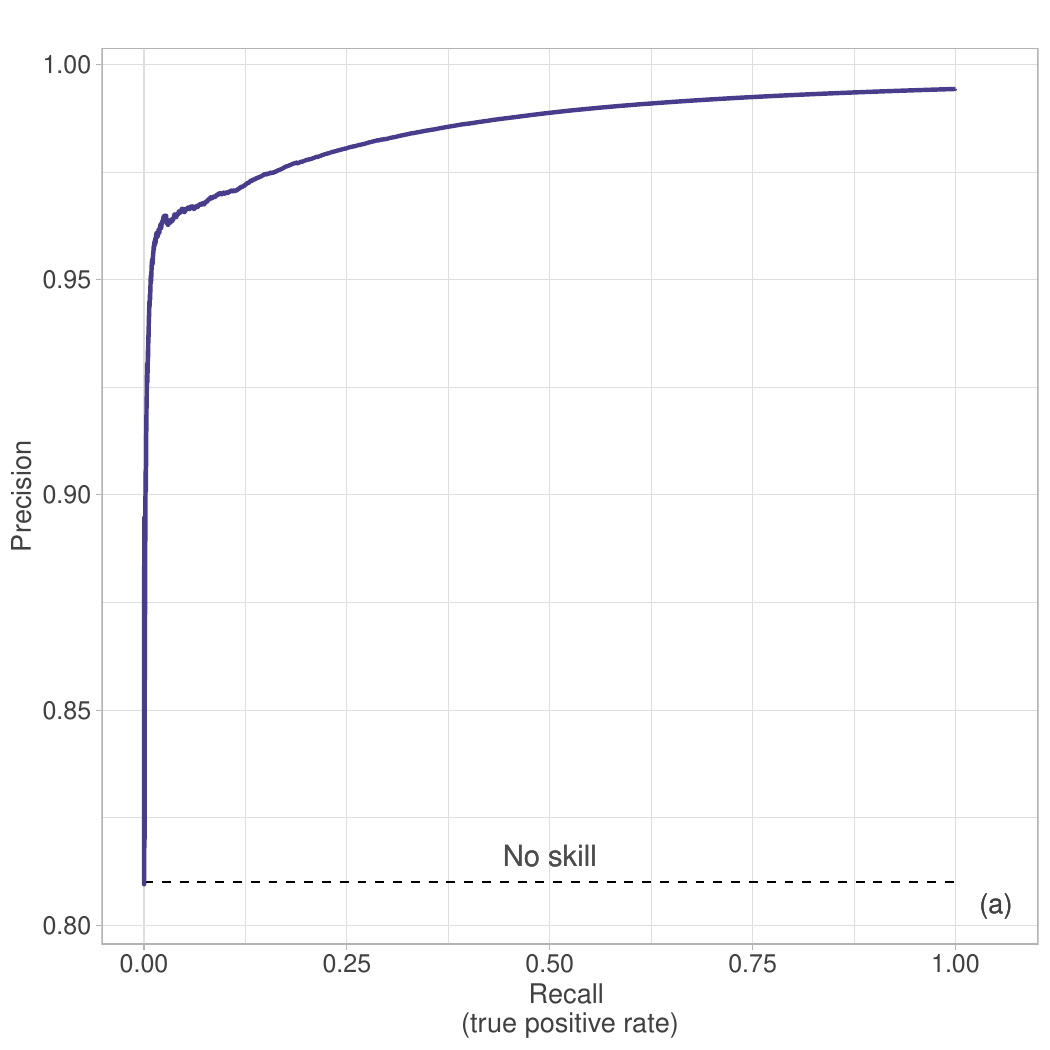}}
\subfloat{\includegraphics[width=0.495\textwidth]{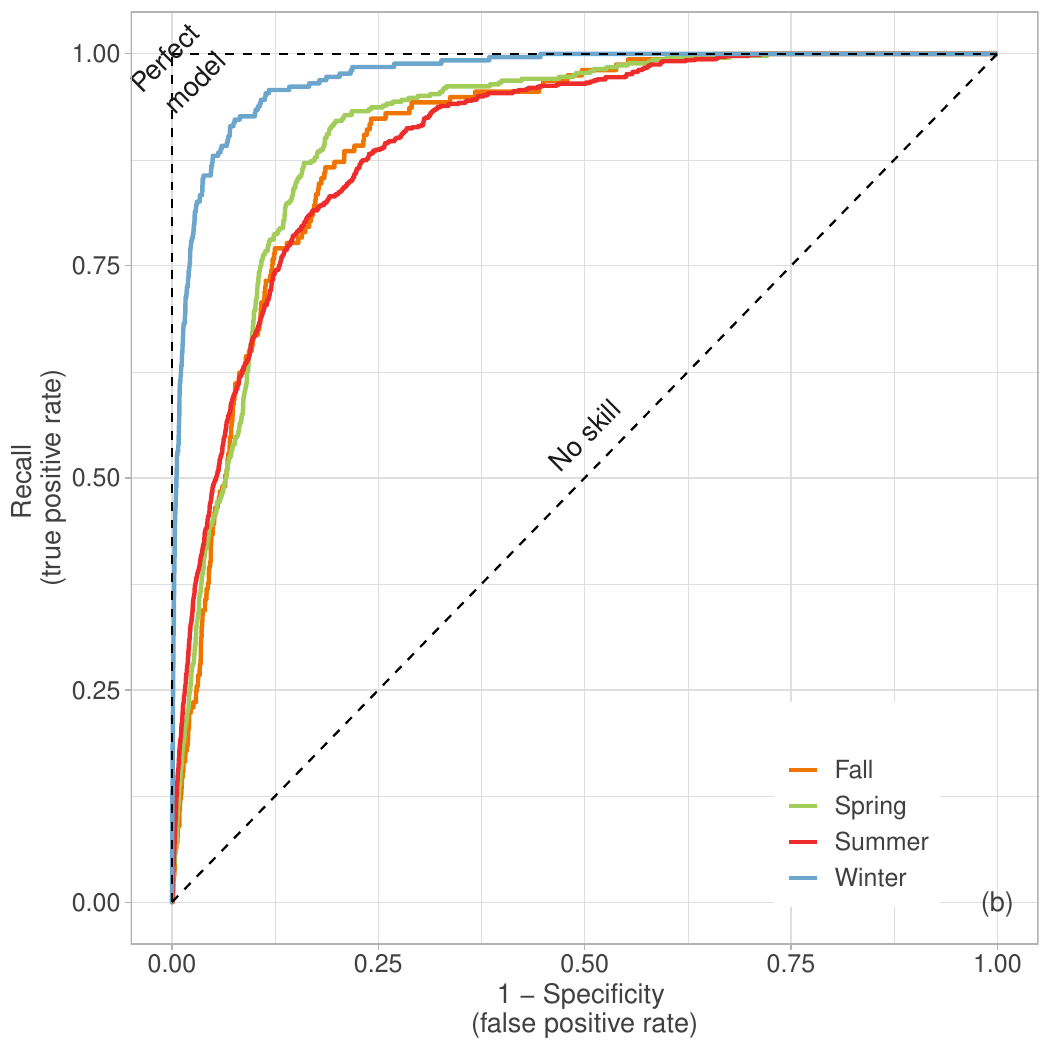}}
\caption{Performance of the random forest models compared to no-skill models. Panel a: precision-recall curve illustrating the trade-off between what proportion of actual UL flashes the model correctly identified (recall), and what proportion of UL flashes predicted by the model actually occurred (precision) for varying cutoff values determining whether UL occurred or not.  Panel b: ROC curves for each season showing the trade-off between the proportion with no UL incorrectly predicted as having UL and how well the models predict UL situations that have actually occurred. The larger the area under the curve in both panels, the better the performance.}
\label{fig:pr_curve}
\end{figure}

UL is rare resulting in a highly imbalanced dataset with a substantially higher fraction of instances where no UL occurs. To evaluate the performance of the Gaisberg Tower-trained random forest models in the study area, two statistical approaches are employed. The basis to understand Fig.~\ref{fig:pr_curve} is to understand the principle of a confusion matrix explaining the differences between true/false positives/negatives (see supporting information). 
The performance results are adjusted to fit the ERA5 grid cell size instead of the original 1~km$2$, which makes it easier to accurately predict lightning at tall objects over time and space. In these adjusted predictions, only the highest predicted conditional probability within each ERA5 grid cell is considered.

Figure~\ref{fig:pr_curve}a shows the precision-recall curve, selected for its ability to handle imbalanced data. In contrast, Figure~\ref{fig:pr_curve}b illustrates the Receiver Operating Characteristic (ROC) curve, a commonly used method for analyzing model classification performance or to compare different models. For both approaches the area under the curve represents the performance, which increases for larger areas.

The precision-recall curve focuses on the positive class, i.e., the UL occurrence and minority in the data set. It evaluates the relationship between the recall or true positive rate, i.e., what proportion of actual UL flashes the model correctly identified, and the precision, i.e., what proportion of UL flashes predicted by the model actually occurred. 
The curve shows how precision and recall change at different cutoff values for distinguishing between UL and no UL. In this case, a precision-recall curve that rises rapidly with increasing recall and levels off slightly in the upper right corner indicates satisfactory model precision, especially in the early stages of recall. The rapid increase in precision at lower recall values demonstrates that the models are accurately identifying UL when it actually occurs, while minimizing the number of actual UL events missed.  Seasonally, the precision-recall curves are almost indistinguishable.

Complementing the precision-recall curve, the ROC curve in Figure~\ref{fig:pr_curve}b shows that the models perform best in winter, as indicated by the blue curve. The ROC curve illustrates the trade-off between how many situations with no UL are incorrectly predicted as having UL and how well the models predict UL situations that have actually occurred. 


\subsection{The larger-scale meteorological influence on the risk of UL}

The random forest model takes advantage of information contained in the 35 meteorological input variables. It also allows to identify the variables containing most information about the occurrence of UL.

\subsubsection{The most influential meteorological variables at the Gaisberg Tower}\label{sec:varimp}
 To calculate the individual impact of each meteorological predictor variable in classifying UL, the values of each predictor variable are randomly shuffled, and the resulting decline in performance is assessed. The larger the decline the more important that variable is.

\begin{figure}
\centering
\includegraphics[width=0.8\textwidth]{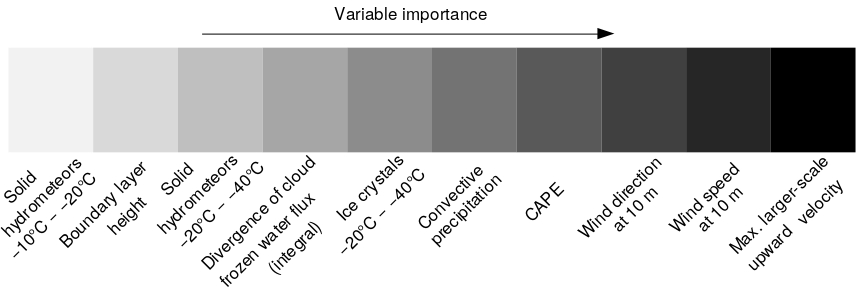}
\caption{Permutation variable importance according to random forests based on balanced proportions of situations with and without UL at the Gaisberg Tower. Importance increases from left to right.}
\label{fig:varimp}
\end{figure}

As evident in the summarized variable importance presented in Fig.~\ref{fig:varimp}, one can deduce that both the wind field and cloud physics-related variables exert most influence on the UL occurrence at the Gaisberg Tower, which is in line with earlier research findings \cite{Stucke2022iclp,Stucke2024}. The top five variables include maximum larger-scale upward velocity, 10 m wind speed, 10 m wind direction, convective available potential energy (CAPE), and convective precipitation. Subsequent analyses will specifically focus on the top three most important variables to enhance our understanding of the modeled risk of UL at tall objects. The maximum larger-scale upward velocity should not be confused with the updrafts associated with the convective processes involved in thunderstorm development. Rather, it is the result of larger-scale processes such as lifting along fronts, synoptic troughs or topography.

\subsubsection{Seasonal analysis of the larger-scale meteorology during lightning at tall objects}\label{sec:meteorology}
\begin{figure}
\centering
\subfloat{\includegraphics[width=.95\textwidth]{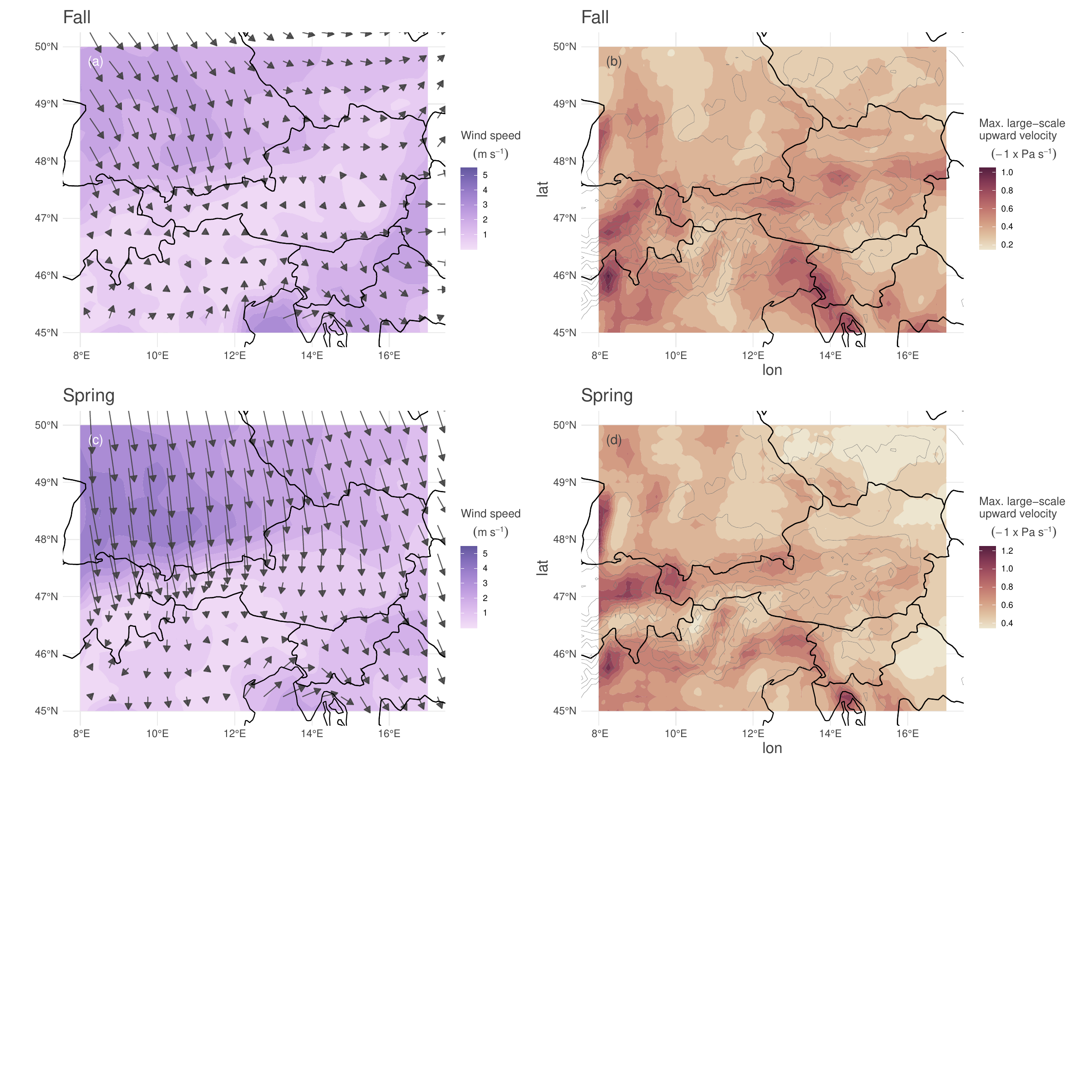}}\vspace{-30ex}
\subfloat{\includegraphics[width=.95\textwidth]{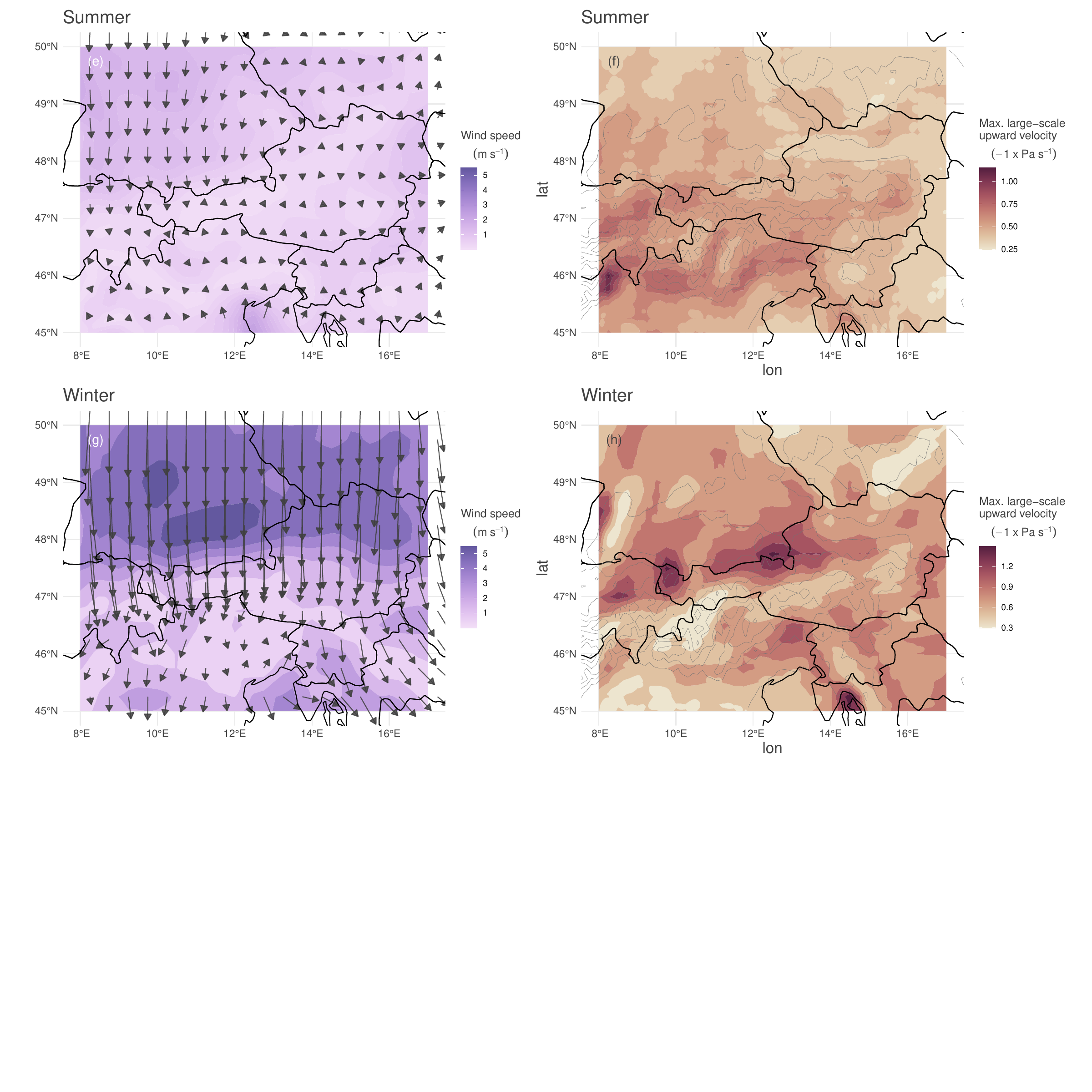}}\vspace{-20ex}
\caption{Seasonal median of the three most influential meteorological variables during LLS-observed lightning at tall objects. Left column: wind speed coded by color and wind direction indicated by arrows (average over $0.5$~$^\circ$~$\times$~$0.5~^\circ$). Right column: Median of the maximum larger-scale upward velocity for each season. Negative values indicate upward motion.}
\label{fig:meteo}
\end{figure}

Each row in Fig.~\ref{fig:meteo} represents a season and shows a distinct meteorological setting prevalent during LLS-observed lightning at tall objects. The panels summarize the median wind speed and wind direction at 10~m (left column) and the median maximum larger-scale upward velocity (right column).

The increased predicted risk in the German subarea as depicted in Fig.~\ref{fig:seas_risk} is associated with northerly and northwesterly near-surface winds in all four seasons. Coupled with hilly terrain, where the winds are deflected upward, this causes enhanced larger-scale upward velocities. Consequently, a relatively high risk of UL is evident throughout the year, with the most significant impact observed in the transitional seasons and winter. 

Similarly, the increased risk associated with complex terrain appears to result from increased maximum upward velocities, likely induced by strong winds impinging the topography and being deflected upward, triggering convection and UL at tall objects. Depending on the prevailing wind direction, increased larger-scale upward velocities are observed either north or south of the eastern Alps (right column). 

Overall, it appears that regions located on the windward side have an increased risk of UL due to comparatively strong near-surface winds and the presence of hills and mountains that deflect the wind upward, creating conditions favorable for UL on tall objects. This is true for the windward side of the northern Alps, which are influenced by strong northerly winds in northern Switzerland, Austria, and the entire German subarea during the transitional seasons and winter. This might also be true for the weak southerly flow, which might influence the risk in western and eastern northern Italy, especially in summer. Conversely, the risk is lower in the central southern Alpine regions of Austria, central southern Switzerland, and central northern Italy.

We propose that especially in winter, and also in spring and fall, processes associated with cyclogenesis, cold front passages, and troughs induce large wind speeds, convective precipitation, and an unstable atmosphere conducive to initiating convection and UL. In contrast, the summer situation might be often characterized by smaller-scale processes and/or strong diurnal heating and solar irradiation, providing conditions for both deep convection initiation and UL at tall objects triggered by nearby DL activity \cite{stuckejgr}. 


\subsubsection{Case study}\label{sec:casestudy}

A case study of the early morning hours (3--6~UTC) of February 21, 2022 demonstrates the performance of the random forests. For simplicity, again only the three most important meteorological variables out of 35 are examined in detail.


The synoptic situation in this case study is dominated by the passage of a cold front, evident from the densely packed isothermes in panel b. The blue line with triangles illustrates the approximate location of the cold front at 6~UTC after having passed through the north-western corner of the study area. The region with high predicted conditional probabilities is characterized by strong near-surface winds originating from the north, peaking in the area where most actual lightning flashes were observed (panel c). Elevation contour lines in panel a indicate elevated terrain, resulting in increased maximum upward velocity when the wind gets deflected. This, in turn, enhances the probability of UL, particularly in the southwesternmost part of Germany, where actual UL flashes have been observed, as indicated by the yellow dots. 

In panel d, a substantial area exceeds a conditional probability value of 0.5, which is the threshold chosen in Fig.~\ref{fig:seas_risk}. The highest predicted probabilities, surpassing 0.8, are concentrated in the German subarea, particularly from western to central southern Germany. Observed lightning at tall objects aligns with the areas of increased risk of UL. However, not all grid cells with elevated probability do experience UL.

\begin{figure}

    \subfloat{\includegraphics[width=0.5\textwidth]{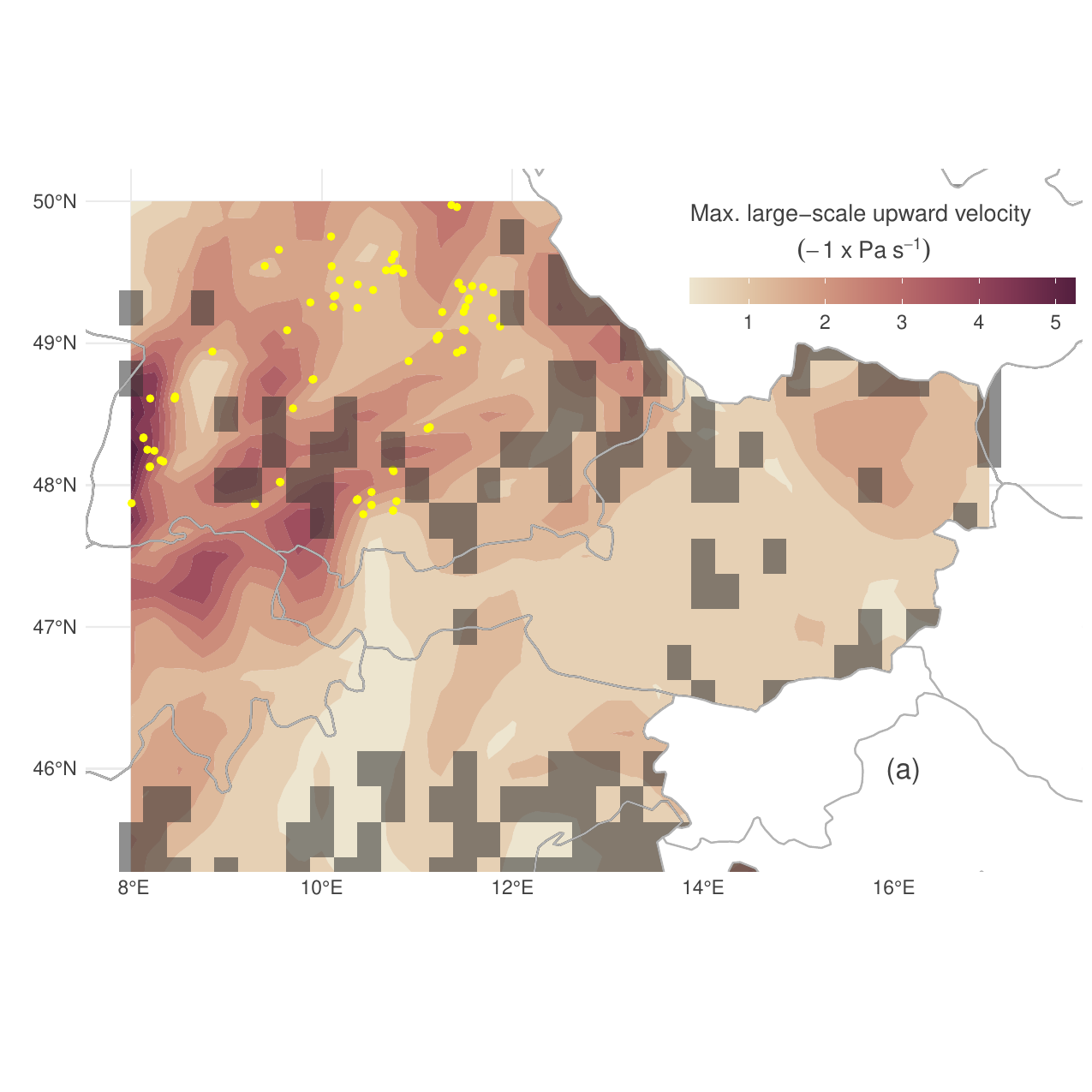}}
    \subfloat{\includegraphics[width=0.45\textwidth]{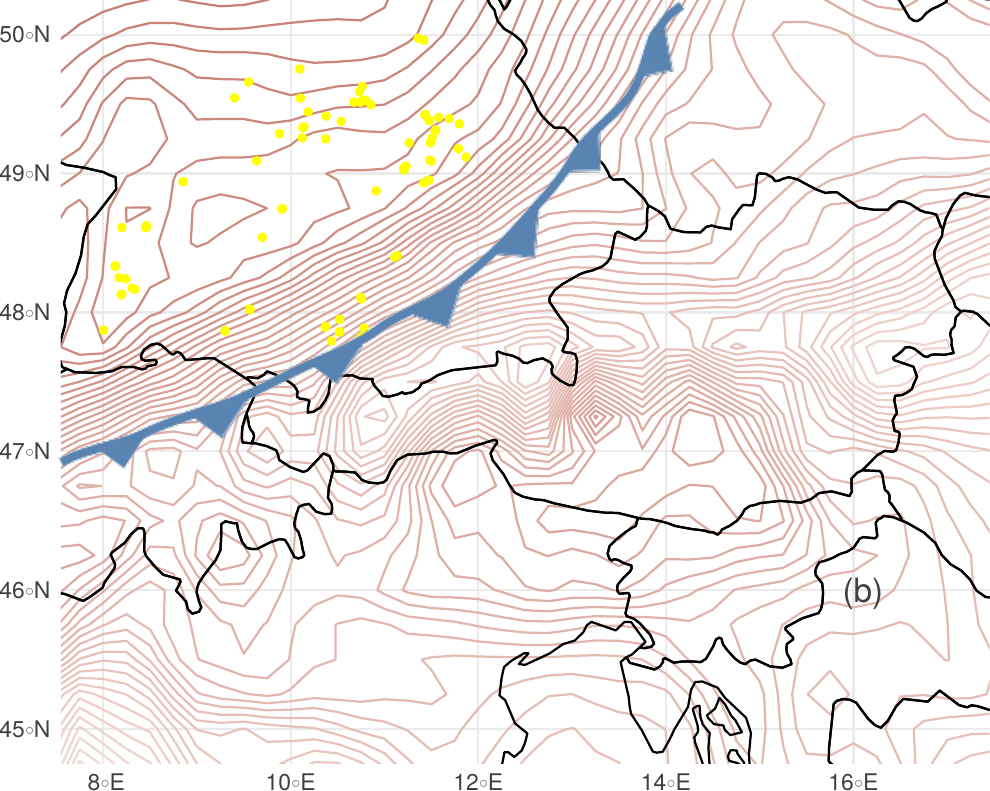}}\hspace{1cm}
   \vspace{-30ex}
    \subfloat{\includegraphics[width=0.48\textwidth]{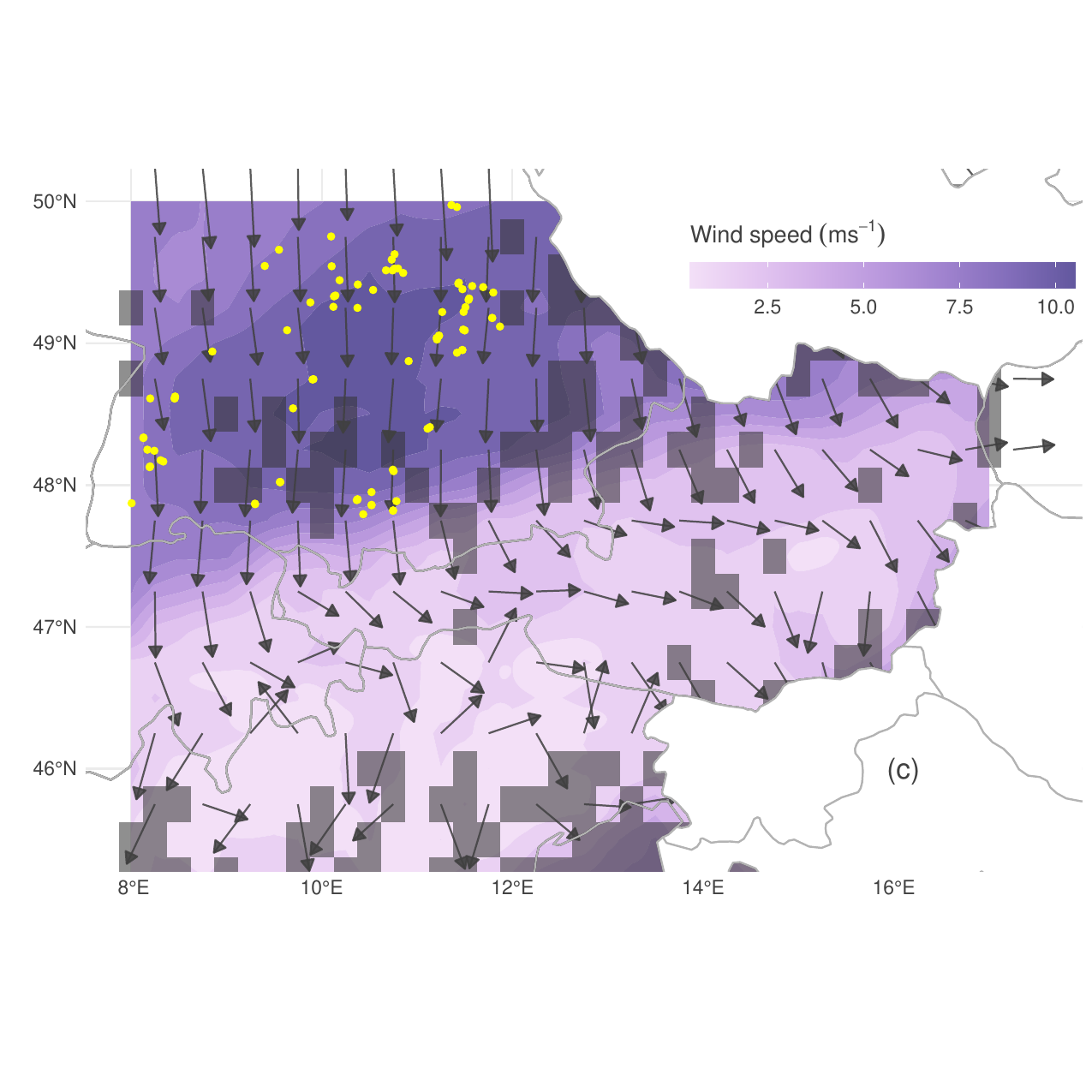}}
    \subfloat{\includegraphics[width=0.48\textwidth]{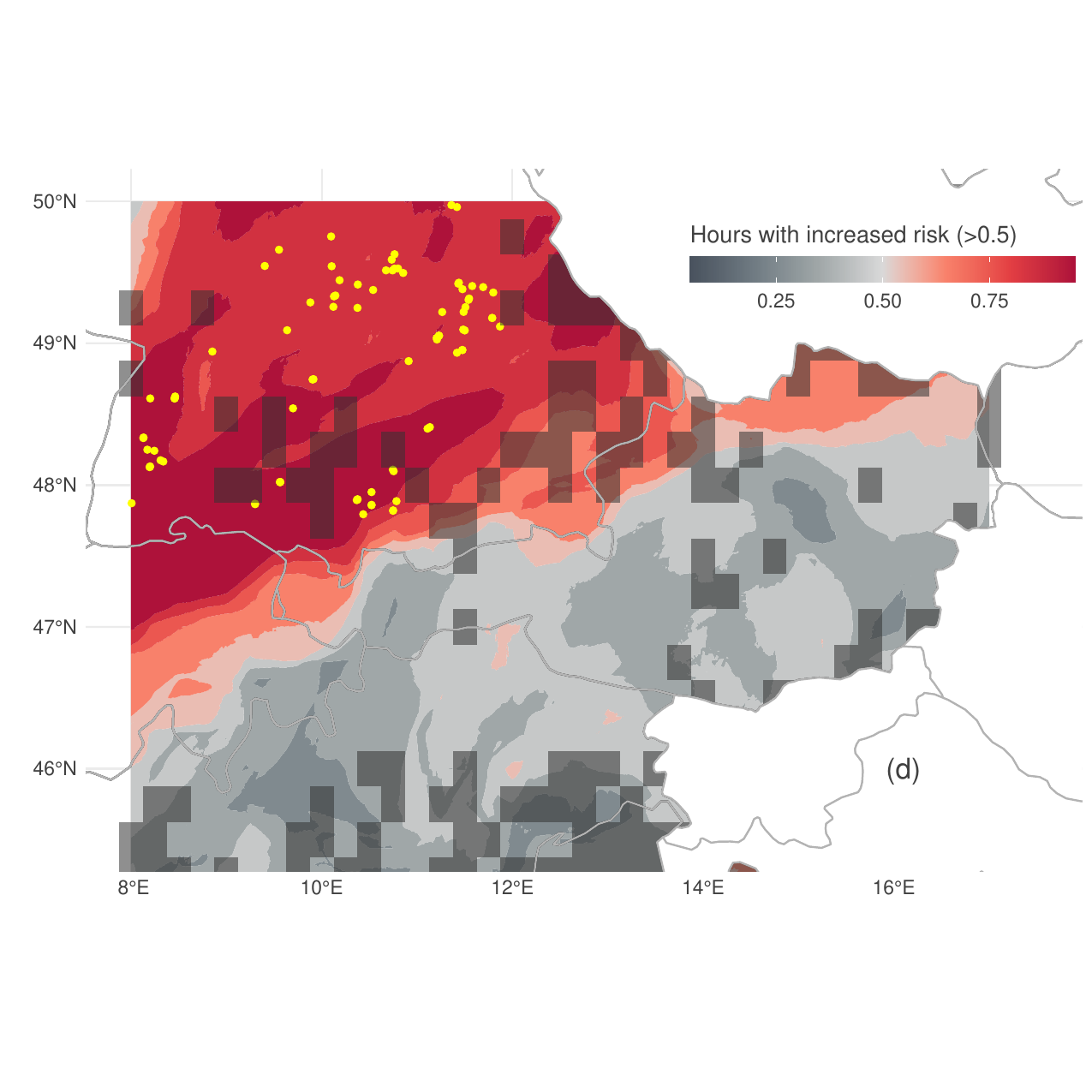}}
 \vspace{25ex}
\caption{Case study from February 21, 2022 between 3~UTC and 6~UTC. Panel a: maximum of the larger-scale upward velocity over verification period. Panel b: Location of 850 hPa isothermes at 6~UTC with the approximate location of the cold front. Panel c: Color areas are maximum of wind speed over verification period, arrows illustrate wind direction at 6~UTC. Panel a: Maximum of predicted  conditional probability over considered verification period. Yellow dots are accumulated LLS-detected flashes at tall structures. Dark gray shaded cells are cells without tall objects.}
\label{fig:case_study}
\end{figure}


\section{Discussion}\label{sec:discussion}

The findings provide clear indications that the seasonal variability in preferred larger-scale meteorological patterns influences the risk of UL at tall objects. Certain regions exhibit higher susceptibility during specific seasons, as also evidenced by observed lightning at tall objects. For instance, in the colder season, the risk is considerably higher north of the Alps. This might be attributed to processes connected to cyclogenesis preferably evolving from north-/north-west to east in the colder season. Conversely, certain areas of northern Italy, particularly the western and eastern parts, where the overall lightning activity is quite high, show a relatively high risk for UL during the summer, in contrast to the lower risk during the colder season. The prevailing favorable meteorological conditions combined with obstructive terrain and elevated effective heights, especially in the hilly regions of southern Germany, may cause the risk to exceed the risk predicted by the random forest models trained on the Gaisberg Tower. 

Although observed lightning at tall objects indicate a reasonable risk assessment, there are naturally discrepancies between the modeled risk and the observation.
The most obvious reason for discrepancies is the fact that the models trained at Gaisberg Tower consider only UL and ignore DL, since the former is almost exclusively observed at Gaisberg Tower. While the models only consider UL, lightning at tall objects used for verification may include both UL and DL, since LLS do not distinguish UL from DL.  
 Consequently, the models may not adequately capture the prevalence of DL at tall objects. This might be less critical in the winter season, which is suggested to be dominated by UL \cite{diendorfer2020,Rachidi2008}. Especially in the late afternoon and evening in summer, the models underestimate the risk of observed lightning at tall objects, while the increased number of observed lightning at tall objects could actually be majorly DL at tall objects and not UL striking the object (see~Fig.~\ref{fig:daily_cycle}).

Another aspect is that successful verification depends on the availability of high quality lightning data. Although the LLS has a high detection efficiency for DL, its efficiency for UL is less than 50\%, which poses a challenge for a reasonable verification of the modeled risk. Although the models exclude ICC\textsubscript{only} UL, both ICC\textsubscript{RS} and especially ICC\textsubscript{Pulse} UL also face limitations in detection efficiency (see also Sect.~\ref{sec:data}).

Other non-meteorological factors may significantly influence the occurrence of UL at wind turbines. Neither topographic characteristics nor varying effective heights can be accounted for in the tower-trained models. As mentioned, the occurrence of UL at tall objects is closely related to the effective height, with both UL and DL possible in the range of approximately 100~m to 500~m. The Gaisberg Tower has a specific effective height of about 270 m according to \cite{Zhou2010} and considerably higher according to \cite{Smorgonskiy2012}. Consequently, the maps in Fig.~\ref{fig:seas_risk} show the risk for objects in this height range. Figure~\ref{fig:effH_statistics}b may be used to adjust it for objects of different heights.

Applying the same algorithm \cite{Zhou2010} to compute the effective height as for all other objects, the effective height of Gaisberg Tower is 270 m. Since it sits on a hill that is approximately 800 m higher than the terrain to the north, its actual effective height likely exceeds 500 m and was determined \cite{Smorgonskiy2012} to range between approximately 300~m to 670~m.
From the results we suggest that the combination of favorable meteorological conditions and increased effective heights, as is especially the case in southern and southwestern Germany and easternmost Austria, could increase the fraction of UL over DL in total lightning at tall objects.

Physical properties of the object may also play a role, for example, the shape of the structure, as well as the rotation of the wind turbine blades may affect the UL risk \cite{Montanya2014}. In addition, wind farms with many turbines can create "hotspots" for lightning due to a significant increase in the electric field \cite{Soula2019}. This would also support the hypothesis that the German subarea, where many wind turbines are located, has the highest proportion of hours in which only lightning at tall objects occurs without any other lightning activity to the ground around the turbine.

Finally, it is often much more important to correctly predict a high risk at the appropriate time, when the event actually occurs, than to overestimate it. The performance analysis and verification have shown that the random forest models trained at Gaisberg Tower are able to reliably and correctly assess this risk, which has the most valuable application also for the wind energy sector.


\section{Conclusions}
This study examines the risk of lightning at tall objects large enough to experience a significant proportion of rare but destructive upward lightning (UL). In recent years, UL has become a major concern for wind turbines as they increasingly suffer from UL. Direct lightning current measurements at the specially instrumented Gaisberg Tower in Austria show that more than half of the UL is not detected by the local Lightning Location System (LLS) due to very specific current waveforms observed in UL making a proper spatio-temporal risk assessment of UL nearly impossible.
Current approaches to assessing lightning risk often overlook crucial meteorological factors, potentially leading to a considerable underestimation of UL risk for wind turbines. This study highlights the necessity of integrating detailed meteorological data into risk assessment to achieve a more reliable understanding of lightning risk at tall wind turbines.

Therefore, this study investigates the larger-scale meteorological role of UL at tall objects and uses direct UL observations at the Gaisberg Tower together with globally available larger-scale meteorological reanalysis data. Random forests, a popular and flexible machine learning technique, distinguish UL from non-UL situations.
 The results show the importance of wind field and cloud physics relevant variables, which is in agreement with previous studies. The three most important variables from a set of 35 distinguishing UL from no-UL situations at Gaisberg are the maximum large-scale upward velocity, wind speed at 10~m, and wind direction at 10~m. Further convective available potential energy and cloud physics  related variables are important.

In a second step, these findings are applied to a study area covering Austria, parts of Italy, Germany and Switzerland. The models trained at the Gaisberg Tower predict the conditional probability of UL within this area at a resolution of 1~km$^2$. 
For verification, all objects large enough to experience UL, i.e., having an effective height of $\geq$~100 m, are considered, and LLS-detected lightning at tall objects in the verification period between 2021 and 2023 within a 100~m radius of each tall object are extracted.
Tall objects are distributed throughout the study area, with maxima in the central-eastern Swiss subarea and eastern Austria. Objects with large effective heights are found in southern, south-western and central Germany, as well as eastern Austria. 

 The highest LLS-observed activity of lightning at tall objects is mainly in the central southern and western German subarea, as well as in the Swiss subarea. Wind turbines are most pronounced in the German subarea and in easternmost Austria. In the German subarea, lightning at tall wind turbines can account for up to 20~\% and more of the total lightning activity within a 10~km radius particularly around wind turbines. In all other subareas the proportion of lightning at tall objects to the total lightning activity 10~km around an object is less than 5~\%.

Evaluating the risk of UL at tall objects from Gaisberg Tower-trained random forest models based only on larger-scale meteorological variables shows that the annual risk is highest in southern Germany as well as northern and eastern Austria and northern Switzerland. Western and eastern northern Italy also have an increased risk of UL.
A seasonal analysis shows that in winter the highest risk is limited to the regions north and east of the eastern Alps, while south of the eastern Alps (eastern and western northern Italy) the risk is also increased in the transition seasons and especially in summer. The analysis of the three main variables shows that the highest predicted probabilities are due to the deflection of strong larger-scale near-surface winds at the topography, leading to an increase in larger-scale upward velocities. In the winter and transition seasons, the wind is predominantly from the north, increasing the risk of UL north of the Alps.
In the warmer seasons and in summer, the increased risk south of the Alps may be due to other influences, such as thermally driven slope winds, valley winds and mountain-plain circulations. Between the high-risk areas of southern Switzerland, central northern Italy and southern parts of Austria, the risk is lower in all seasons.
The diurnal cycle of the modeled risk varies seasonally. While the transitional seasons show a prominent peak in the afternoon, summer and winter show two prominent peaks. The highest risk in summer is in the late afternoon and evening, while the highest risk in winter is in the late evening and night.

A comparison with LLS-observed lightning at tall objects shows a qualitatively good agreement with increased or decreased risk. While the areas of increased risk are much larger than areas with observed lightning at tall objects (UL is a very rare phenomenon), the performance of the models to correctly predict high risk of UL when lightning has actually occurred at a tall object is good throughout the year. The precision of the predictions is highest in winter.


\section*{acknowledgements}
We acknowledge the funding of this work by 
the Austrian Climate Research Program - Implementation.
The computational results presented have been achieved in part using the Vienna Scientific Cluster (VSC).

\section*{conflict of interest}
The authors declare no competing interests.

\section*{data availability}
ERA5 data are freely available at the Copernicus Climate Change Service (C3S) Climate Data Store \cite{Hersbach2020}. The results contain modified Copernicus Climate Change Service information (2020). Neither the European Commission nor ECMWF is responsible any use that may be made of the Copernicus information or data it con-
tains. EUCLID data and ground truth lightning current measurements from the Gaisberg Tower are available only on request. For more details contact Wolfgang Schulz.
\clearpage
\bibliography{References}
\clearpage
\section*{Supporting Information}

\subsection*{Estimation of the effective height}
The effective height is computed following \cite{Zhou2010} by assuming a hemispherical mountain:

\begin{equation}
\begin{aligned}
 & \quad H_{\text{eff}} = \left(5.87 \times 10^{-3} + 2.04 \times 10^{-6} \times E_g\right)^{-2.04 \times 10^{-6}} \\
 \end{aligned}
\end{equation}

using:

\begin{equation}
\begin{aligned}
& \quad E_g = \frac{U_{\text{lc}} + 73400}{(h + a)(1 - \frac{a^3}{(h + a)^3})} \\
\\
& \quad U_{\text{lc}} = \frac{1556 \times 10^3}{1 + \frac{7.78}{R}} \\
\\
& \quad R = \frac{2(h+a)}{1 + \left(\frac{r_1}{r_2-a^2}\right) - \left(\frac{r_1}{r_2+a^2}\right)} \\
\\
& \quad r_1 = 2a(h+a); \quad r_2 = (h+a)^2\\
\\
\end{aligned}
\end{equation}

where $H_{eff}$ (m) is the effective height and $h$ (m) is the actual height of the object. $U_{lc}$ (kV) is the continuous leader inception potential due to the cloud charges, $R$ (m) is a geometrical parameter, $a$ (m) is the mountain height, which in the current study is taken to be the difference between the 1~km$^2$ mean elevation and the elevation at which the object is located to also account for the surrounding terrain.  $E_g$ (kV/m) is the ambient uniform electric field. For more details see \cite{Zhou2010}.

\subsection*{Example of a decision tree}\label{sec:dectree}

\begin{figure}
\centering
\includegraphics[width=0.7\textwidth]{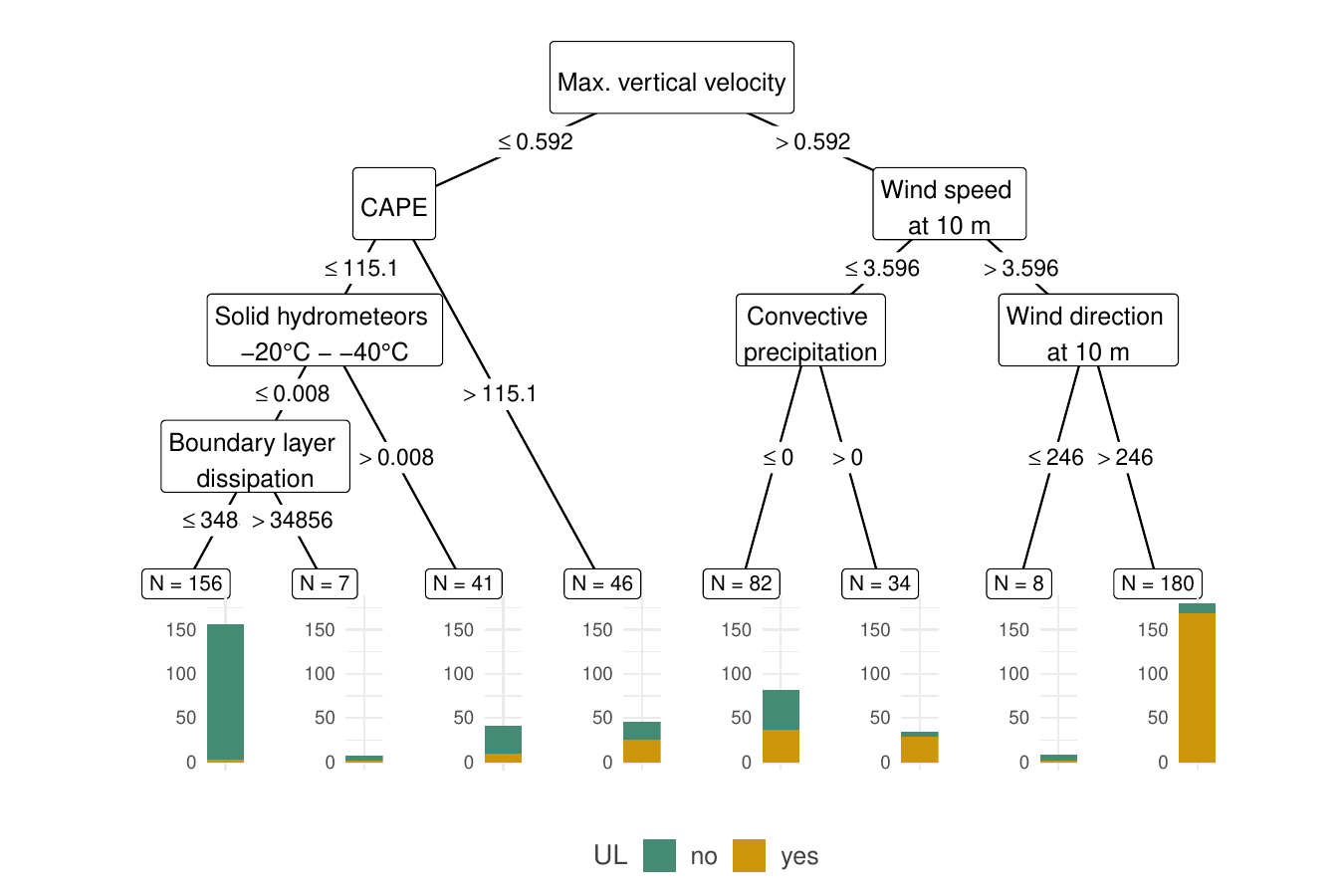}
\caption{Example of a decision tree. Meteorological variables in the nodes are splitted according to the split points (numbers at the solid lines). Terminal nodes (bars) give the decision. The number of observations included in the decision pars is given above the terminal nodes as $N$.}
\label{fig:ctree}
\end{figure}

Figure \ref{fig:ctree} shows the structure of a single decision tree. It shows several nodes, each associated with specific split variables. Initially, the maximum large-scale upward velocity serves as the primary split variable. Thresholds between nodes indicate where the split variable is splitted for optimal performance. Following a single UL observation along the path determined by these thresholds leads to a terminal node, represented by the bottom bars. The colors of these bars indicate the number of observations assigned to each terminal node, indicating UL or no UL prediction.

\subsection*{Understanding a confusion matrix}

\begin{tabular}{l|l|c|c|c}
\multicolumn{2}{c}{} & \multicolumn{2}{c}{Actual} & \\
\cline{3-4}
\multicolumn{2}{c|}{} & Positive & Negative \\
\cline{2-4}
\multirow{2}{*}{Predicted} & Positive & True positive & False positive \\
\cline{2-4}
& Negative & False negative & True negative \\
\cline{2-4}
\multicolumn{1}{c}{} & \multicolumn{1}{c}{Total} & \multicolumn{1}{c}{$a+c$} & \multicolumn{1}{c}{$b+d$} & \multicolumn{1}{c}{$N$}\\
\end{tabular}

\vspace{1cm}

A true positive rate is the proportion of true positive divided by the sum of true positives and false negatives. The false positive rate on the other hand is the proportion of false positives divided by the sum of true positives and false positives.

\subsection*{List of variables included in the random forest models}
\begin{table}
    \caption{Table of larger-scale variables taken from ERA5 and variables derived from ERA5. The derived variables are suggested to be potentially important in the charging process of a thundercloud or for the development of convection.}
    \setlength\extrarowheight{-6pt}
    \vspace{1.5cm}
    \begin{tabular}{ll|ll}
        \toprule
        \textbf{Variable} & \textbf{Unit} & \textbf{Variable} & \textbf{Unit} \\
        \midrule
        Cloud base height above ground & m agl & Convective precipitation (rain + snow) & m \\
        Large scale precipitation & m & Cloud size & m \\
        Maximum precipitation rate (rain + snow) & kg m$^{-2}$ s$^{-1}$ & Ice crystals (total column, tciw) & kg m$^{-2}$ \\
        Solid hydrometeors (total column, tcsw) & kg m$^{-2}$ & Supercooled liquid water (total column, tcslw) & kg m$^{-2}$ \\
        Water vapor (total column) & kg m$^{-2}$ & Integral of cloud frozen water flux divergence & kg m$^{-2}$ s$^{-1}$ \\
        Vertical transport of liquids around $-10$ $^\circ$C & kg Pa s$^{-1}$ & Ice crystals ($-10$ $^\circ$C - $-20$ $^\circ$C) & kg m$^{-2}$ \\
        Ice crystals ($-20$ $^\circ$C - $-40$ $^\circ$C) & kg m$^{-2}$ & Cloud water droplets ($-10$ $^\circ$C - $-20$ $^\circ$C) & kg m$^{-2}$ \\
        Solid hydrometeors ($-10$ $^\circ$C - $-20$ $^\circ$C) & kg m$^{-2}$ & Solid hydrometeors ($-20$ $^\circ$C - $-40$ $^\circ$C) & kg m$^{-2}$ \\
        Solids (cswc + ciwc) around $-10$ $^\circ$C & kg m$^{-2}$ & Liquids (clwc + crwc) around $-10$ $^\circ$C & kg m$^{-2}$ \\
        2 m dew point temperature & K & Mean vertically integrated moisture convergence & kg m$^{-2}$ s$^{-1}$ \\
        Water vapor ($-10$ $^\circ$C - $-20$ $^\circ$C) & kg m$^{-2}$ & Boundary layer height & m \\
        Surface latent heat flux & J m$^{-2}$ & Surface sensible heat flux & J m$^{-2}$ \\
        Downward surface solar radiation & J m$^{-2}$ & Convective available potential energy & J kg$^{-1}$ \\
        Convective inhibition present & binary & Mean sea level pressure & Pa \\
        Height of $-10$ $^\circ$C isotherm & m agl & Boundary layer dissipation & J m$^{-2}$ \\
        Maximum larger-scale upward velocity & Pa s$^{-1}$ & Total cloud shear & m s$^{-1}$ \\
        Wind speed at 10 m & m s$^{-1}$ & Wind direction at 10 m & $^\circ$ \\
        Shear between 10 m and cloud base & m s$^{-1}$ \\
        \bottomrule
    \end{tabular}
    \label{tab:era5_names}
\end{table}

\end{document}